\documentclass[showpacs,floatfix,eqsecnum,aps,nofootinbib, amssymb]{revtex4}

\begin{document}
\title{Unification via intermediate symmetry breaking scales with the quartification gauge group.}  
\author{Alison Demaria}\email{a.demaria@physics.unimelb.edu.au} \affiliation{School
of Physics, Research Centre for High Energy Physics, The University of
Melbourne, Victoria 3010, Australia} \author{Catherine I. Low}\email{c.low@physics.unimelb.edu.au} \affiliation{School
of Physics, Research Centre for High Energy Physics, The University of
Melbourne, Victoria 3010, Australia} \author{Raymond
R. Volkas}\email{r.volkas@physics.unimelb.edu.au} \affiliation{School
of Physics, Research Centre for High Energy Physics, The University of
Melbourne, Victoria 3010, Australia}

\begin{abstract}
The idea of quark-lepton universality at high energies has been introduced  
as a natural extension to the standard model. 
This is achieved by endowing leptons with new degrees of freedom -- leptonic colour, an 
analogue of the familiar quark colour. Grand and partially unified models which 
utilise this new gauge symmetry $SU(3)_\ell$
have been proposed in the context of the quartification gauge group $SU(3)^4$.  
Phenomenologically successful gauge coupling constant unification without supersymmetry has
been demonstrated for cases where the symmetry breaking 
leaves a residual $SU(2)_\ell$ unbroken. Though attractive, these schemes either incorporate 
ad hoc discrete symmetries and non-renormalisable mass terms, or achieve only partial unification. 
We show that grand unified models can be constructed where the 
quartification group can be broken fully [i.e.\ no residual $SU(2)_\ell$]
to the standard model 
gauge group without requiring additional discrete symmetries or higher dimension operators. 
These models also automatically have suppressed nonzero neutrino masses.
We perform a systematic analysis of 
the renormalisation-group equations for all possible symmetry breaking routes from 
$SU(3)^4 \rightarrow SU(3)_q \otimes SU(2)_L \otimes U(1)_Y$. 
This analysis indicates that gauge coupling unification can be achieved for 
several different symmetry breaking patterns and we outline the requirements that each 
gives on the unification scale. We also show that the unification scenarios of those 
models which leave a residual $SU(2)_\ell$ symmetry are not unique. 
In both symmetry breaking cases, some of the scenarios require new physics at the $TeV$ scale,
while others do not allow for new $TeV$ phenomenology in the fermionic sector. 
\end{abstract}

\maketitle

\section{Introduction}
Grand unified theories (GUTs) are an important class of extensions to standard model (SM) physics,
 with most theories attempting to unify the strong and electroweak interactions within 
 the framework of a single, larger gauge symmetry $G$. The simple groups $SU(5)$~\cite{su5} 
and $SO(10)$~\cite{so10}, 
which may be derived from a possible underlying $E_6$~\cite{e6}, 
have been the common groups of interest. $SU(5)$ is the smallest group with 
complex representations 
 that can accommodate the SM gauge structure, with the fermions placed in the 
$\mathbf{1 \oplus \overline{5} \oplus 10}$
 representations. As both quarks and leptons are contained in both the 
$\mathbf{\overline{5}}$ and the 
 $\mathbf{10}$, gauge-mediated quark-lepton transformations exist, giving rise to 
baryon number violation. 
 Similarly, $SO(10)$ can house an entire generation of fermions, including 
the right-handed neutrino, in a 
 single $\mathbf{16}$. This enumeration of fermions provides great simplicity 
but also places unrealistic bounds on 
 proton stability once a unification scale is identified. 

Additionally, these groups are plagued by a lack of 
phenomenologically successful coupling constant unification. 
As these unified theories
 are based upon a simple group $G$, a single gauge coupling constant describes 
the strength of all 
gauge interactions. The three coupling parameters of the low energy SM 
field theory need 
to separately evolve as a function of energy until concordance at some possible
 unification energy scale 
results and we have only one effective coupling constant~\cite{gqw}.
Whether unification of the gauge coupling constants can be achieved represents 
a crucial test for the feasibility of a GUT. 
The running of the coupling constants in $SU(5)$ and $SO(10)$ theories fail to 
satisfy this criterion unless appropriate new physics at an intermediate scale, such as 
supersymmetry, is invoked. 
This motivates the application of product groups 
$G \otimes G \otimes ...$, 
augmented by a discrete symmetry permuting the $G$ factors, as an alternative class of 
unified theories. These models need not have gauge 
boson mediated proton decay as the quarks and leptons are often in 
separate representations, however even if so, proton instability 
can still orginate through Higgs-fermion Yukawa interactions.

The smallest such group consistent with SM phenomenology at low energy is 
trinification, based on $SU(3)_q \otimes SU(3)_L \otimes SU(3)_R$, which has shown 
promise both within and independent of a supersymmetric context. It can also 
be obtained from an $E_6$ theory and as a result has been studied 
extensively \cite{trinification,Willenbrock}. 

A natural extension to trinification is quartification.\footnote{See also Ref.~\cite{carone1}
for an extension of trinification to $3N$ $SU(3)$ factors, where $N$ is a positive integer.}
Quartification 
was first proposed by Volkas and 
Joshi~\cite{Ray} and then independently revisited by Babu, Ma and 
Willenbrock~\cite{BMW}. 
These theories represent 
an implementation of the idea of quark-lepton universality at high energies 
postulated by Foot and Lew~\cite{Foot}. 
In the low energy world described by the SM, there are significant disparaties between the quarks and 
leptons. They have different electric charges, and the quarks are confined by 
colour interactions whereas the leptons are not. Extended models employing discrete quark-lepton
symmetry~\cite{Foot} allow quarks and leptons to become indistinguishable
above energy scales as low as a few $TeV$. To achieve this, one must introduce new degrees of 
freedom for the leptons which 
are embodied in a separate gauge group $SU(3)_{\ell}$. With this gauge 
group supplementing the familiar quark 
colour $SU(3)_q$ group, a discrete exchange symmetry between the quark and leptons can be imposed.
The quark-lepton indistinguishability afforded by this scheme does not require
gauge coupling constant unification.
Quartification is the simplest known way to extend such models to also
provide for full coupling constant unification.  

The gauge group of these theories is $SU(3)^4$ with an anomaly-free fermion 
assignment. 
The symmetry breaking is accomplished with Higgs multiplets in a certain $36$-dimensional
representation of $SU(3)^4$, and in Refs.~\cite{Ray} and \cite{BMW}
the symmetry is broken down to $SU(3)_q \otimes SU(2)_L \otimes U(1)_Y \otimes SU(2)_{\ell}$. 
In the low-energy limit, 
there is an $SU(2)_\ell$ remnant of leptonic colour left unbroken. Each standard lepton has heavy exotic
partners, which following tradition we call ``liptons'', in an $SU(2)_\ell$ doublet.  If the liptons are heavier
than a few 100 GeV, then the existence of the new unbroken gauge symmetry $SU(2)_\ell$ is hidden,
though potentially to be found at Large Hadron Collider energies. 

The original proposal of Ref.~\cite{Ray} succeeds only in partial unification~\cite{ps}, with 
two independent gauge coupling 
constants at the unification scale. The model of Babu et al.~\cite{BMW} is a variant
that achieves full unification, with 
the running coupling constants meeting at around $4 \times 10^{11}\ GeV$
without invoking 
supersymmetry. Vital to this result is the presence of the liptons at the $TeV$ scale,
to help ensure appropriate running for the coupling constants, assisted by a second
light Higgs doublet.
In non-quartified Foot-Lew type models~\cite{Foot}, $TeV$-scale liptons are a natural
possibility.  In quartification, however, the default situation sees the liptons
acquiring unification-scale masses.  This is precisely why the original paper~\cite{Ray}
proposed partial unification only.  By contrast, Babu et al.~\cite{BMW} impose an
additional discrete symmetry for the sole purpose of avoiding ultra-large lipton masses.
While it is of course technically natural, the additional discrete symmetry is in other
respects very much an afterthought. It is compounded by the 
resolutions employed (non-renormalisable operators) to obtain realistic $TeV$-scale 
lipton masses.  There are also issues with neutrino mass generation.
This scheme is reviewed more fully in Sec.~\ref{cha:BMW}.

The purpose of this paper is to show how to achieve full unification through quartification 
without the imposition of the additional discrete symmetry. In fact, 
we demonstrate that the unification scheme of Ref.~\cite{BMW} is not unique, and can be obtained 
without restricting 
Yukawa couplings simply by introducing intermediate stages in the symmetry breaking. 
Importantly, we show that 
heavy exotic liptons need not spoil unification. 
In Sec.~\ref{cha:masses} 
we describe the matter content of our models and state any assumptions 
associated with the Higgs sector. 
We outline in Sec.~\ref{cha:SU2l} all possible choices of symmetry 
breaking which leave a residual 
$SU(2)_{\ell}$, while in the subsequent subsection we break this symmetry 
entirely~\cite{lbroken} and 
resolve the neutrino mass issues.
We systematically solve the renormalisation-group equations for each, 
showing the choices that give rise to successful
unification. In Sec.~\ref{cha:phenomenology} we comment on the low-energy 
phenomenology in those scenarios 
in which unification is possible. We conclude in Sec.~\ref{cha:conclusion}.

\section{The quartification model} \label{cha:BMW}

The quartification gauge group is 
\begin{equation}
\centering
G_4 = SU(3)_q \otimes SU(3)_L \otimes SU(3)_{\ell} \otimes SU(3)_R.
\label{eqn:quart}
\end{equation} 
A $Z_4$ symmetry which cyclicly permutes the gauge groups as per $q \rightarrow L \rightarrow 
\ell \rightarrow R \rightarrow q$ is imposed, ensuring a single gauge coupling constant $g_4$. 
The fermions are contained within a left-handed $\mathbf{36}$ of Eq.~\ref{eqn:quart},
\begin{eqnarray}
\mathbf{36} & = &\mathbf{(3,\overline{3},1,1)+(1,3,\overline{3},1,)+(1,1,3,\overline{3})+(\overline{3},1,1,3)}, \\
& \equiv & \qquad q \qquad \oplus \qquad \ell \qquad \oplus \qquad {\ell}^c \qquad \oplus \qquad q^c, 
\end{eqnarray}
where $q (\ell)$ denotes the left-handed quarks (leptons) and $q^c ({\ell}^c )$ the left-handed 
anti-quarks(leptons). Under $G_4$, these have the transformations 
\begin{equation}
\centering
q \rightarrow U_q \, q \, U_L^\dagger, \qquad q \rightarrow U_R\, 
q^c\, U_q^\dagger,\qquad \ell \rightarrow U_L\, \ell \,U_\ell^\dagger, \qquad 
\ell^c \rightarrow U_\ell \, \ell^c \, U_R^\dagger,
\label{eqn:fermtransforms}
\end{equation} 
where $U_{q, L, \ell, R} \in SU(3)_{q, L, \ell, R}$, and the 
multiplets are represented by $3 \times 3 $ matrices:
\begin{eqnarray}\label{eqn:fermions}
\centering
q &\sim& (\mathbf{3,\overline{3},1,1})=
\left(\begin{array}{ccc} 
d & u & h \\ 
d & u & h \\  
d & u & h \end{array} \right),\qquad
q^c \sim  (\mathbf{\overline{3},1,1,3})=
\left(\begin{array}{ccc} 
d^c & d^c & d^c \\  
u^c & u^c & u^c \\
h^c & h^c & h^c \end{array} \right),
\\ \nonumber
\ell&\sim&(\mathbf{1,3,\overline{3},1})=
\left(\begin{array}{ccc}x_1&x_2&\nu\\
y_1&y_2&e\\
z_1&z_2&N \end{array} \right),\qquad
\ell^c \sim  (\mathbf{1,1,3,\overline{3}})=
\left(\begin{array}{ccc}x_1^c&y_1^c&z_1^c\\
x_2^c&y_2^c&z_2^c\\
\nu^c&e^c&N^c\end{array}\right).
\end{eqnarray} 
Note the existence of exotic particles as a necessary ingredient in 
the representations.
Defining the generator of electric charge $Q$ as 
\begin{equation}
\centering
Q = I_{3L} - \frac{Y_L}{2} - \frac{Y_{\ell}}{2} + I_{3R} - \frac{Y_R}{2},
\end{equation}
we identify $N, N^c$ as neutral particles, $h(h^c)$ as charge $Q=-\frac{2}{3}(\frac{2}{3})$ exotic quarks
and the liptons $(x,y,z)$ to have charges $(\frac{1}{2},-\frac{1}{2},\frac{1}{2})$.

In Babu et al.'s model~\cite{BMW}, the Higgs fields are contained in two 
different $\mathbf{36}$'s, and shall be denoted
\begin{eqnarray}
\centering
\Phi_a & \sim (\mathbf{1, \overline{3},1,3}), \qquad 
\Phi_b \sim (\mathbf{3,1, \overline{3},1}), \qquad 
\Phi_c \sim (\mathbf{1,3,1, \overline{3}}), \qquad 
\Phi_d \sim (\mathbf{\overline{3},1,3,1}), \nonumber \\
 \Phi_\ell & \sim (\mathbf{1, 3, \overline{3},1}), \qquad 
\Phi_{\ell^c} \sim (\mathbf{1, 1,3, \overline{3}}), \qquad 
\Phi_{q^c} \sim (\mathbf{\overline{3},1,1,3}), \qquad 
\Phi_{q} \sim (\mathbf{3, \overline{3},1,1})
\label{eqn:higgstrans}
\end{eqnarray}
in this paper (Ref.~\cite{BMW} used different subscripts).  Note that $\Phi_a \sim \Phi_c^{\dagger}$; this
effective replication of a Higgs multiplet is achieved as a natural consequence of the $Z_4$
symmetry.
These fields, which comprise two sets of multiplets that are closed under the $Z_4$, 
are sufficient to break the quartification symmetry 
and generate realistic fermion masses and mixings. The VEV pattern is given by 
\begin{eqnarray}
\langle \Phi_{a} \rangle &\sim \langle \Phi_{c}^\dagger \rangle \sim \left(\begin{array}{ccc}
u&0&u\\
0&u&0\\
v&0&v\end{array}\right),\qquad
\langle \Phi_\ell \rangle \sim \left(\begin{array}{ccc}
0&0&u\\
0&0&0\\
0&0&v\end{array}\right),\qquad
\langle \Phi_{\ell^c} \rangle \sim \left(\begin{array}{ccc}
0&0&0\\
0&0&0\\
v&0&v\end{array}\right),\\
\langle \Phi_b \rangle &= \langle \Phi_d \rangle = \langle \Phi_{q^c} \rangle = 
\langle \Phi_{q} \rangle = 
\left(\begin{array}{ccc}0&0&0\\0&0&0\\0&0&0\end{array}\right),\qquad \qquad \qquad \qquad \qquad \qquad \qquad \qquad
\end{eqnarray}
where the $u$'s and $v$'s are at the electroweak and unification scales, respectively. 
This VEV structure induces the strong symmetry breaking 
$SU(3)_q \otimes SU(3)_L \otimes SU(3)_{\ell} \otimes SU(3)_R \rightarrow 
SU(3)_q \otimes SU(2)_L \otimes SU(2)_{\ell} \otimes U(1)_Y$  
in a single step, and then instigates electroweak symmetry breaking. 
To do this, a delicate 
hierarchy within some of these multiplets must exist which necessitates unnatural fine-tuning. (This
unwelcome feature also occurs in trinification models and in the new quartification schemes
we propose below.)
 
The coupling to the fermions is described by the $Z_4$-invariant 
Lagrangian\footnote{The notation $\ell \ell^c \Phi_a$ means $\overline{\ell}_R \ell_L \Phi_a$, etc.}
\begin{equation}\label{eqn:lag}
\centering
\mathcal{L} = Y_1 Tr \left( \ell \ell^c \Phi_a + \ell^c q^c \Phi_b + q q^c \Phi_c + q \ell \Phi_d \right) 
+ Y_2 Tr \left( \ell \ell^c \Phi_c^{\dagger} + \ell^c q^c \Phi_d^{\dagger} + q q^c \Phi_a^{\dagger} 
+ q \ell \Phi_b^{\dagger} \right) + H.c. 
\end{equation}
The theory also admits couplings of the type $\lambda \ell \ell \Phi_\ell$ and cyclic permutations. 
These terms, however, give GUT scale masses to the liptons
$x_1$, $x_2$, $y_1$, $y_2$, $x_1^c$, $x_2^c$, $y_1^c$ and $y_2^c$, 
and, according to~\cite{BMW}, it is essential that these particles remain light for 
gauge coupling constant unification. An additional 
$Z_4^{'}$ symmetry defined by 
\begin{equation}
\centering
(q, \ell, q^c, {\ell}^c) \rightarrow i ( q, \ell, q^c, {\ell}^c), \qquad 
\Phi_{a-d} \rightarrow -\Phi_{a-d} \qquad \textrm{and} 
\qquad \Phi_{\ell,\ell^c,q,q^c} \rightarrow \Phi_{\ell,\ell^c,q,q^c} 
\label{eqn:z4'}
\end{equation}
is then imposed to forbid these terms, 
reducing the Yukawa Lagrangian exclusively to Eq.~\ref{eqn:lag}.
As a consequence of this $Z_4'$, after symmetry breaking, the massless particle 
spectrum contains the $x,x^c$ and $y,y^c$ liptons in 
addition to the minimal SM particles. This massless spectrum, which contains two SM Higgs doublets,  
is sufficient to unify the coupling constants within experimental error at approximately 
$4 \times 10^{11} GeV$.

If the discrete symmetry of Eq.~\ref{eqn:z4'} was not imposed, then the intersection of the three 
SM coupling constants would not occur. A less nice aspect of this scheme is that 
there is no natural origin for this symmetry -- why, on theoretical grounds, should 
certain Yukawa terms be forbidden and not others? This unattractive feature is exacerbated by 
the need to introduce non-renormalisable terms of the form $\epsilon_{jkl}\epsilon_{mnp} \ell^{jm} 
\ell^{kn} ( \Phi_a^{\dagger} \Phi_{\ell^c}^{\dagger} ) ^{lp}$   
and $\epsilon_{jkl}\epsilon_{mnp} (\ell^c)^{jm} 
(\ell^c)^{kn} ( \Phi_{\ell}^{\dagger} \Phi_c )^{lp}$. These terms must exist to give $TeV$ scale 
masses to the liptons, otherwise their mass terms would be indistinguishable from the 
ordinary leptons at the electroweak level.  Although the proton decay mediated by these terms is predicted 
at a realistic rate, from a model-building point of view they are a little ad hoc. 
Babu et al.\ then introduce another ingredient
to resolve the issue of neutrino mass. In the bare model, neutrinos naturally acquire 
Dirac masses of the same order as the charged fermions. 
This is circumvented by the addition of a Higgs singlet whose 
coupling with the right-handed neutrino induces the seesaw mechanism. 
We comment later on how neutrinos naturally acquire light masses without a  
Higgs singlet when $SU(2)_{\ell}$ is broken.  All in all, the very pleasing gauge unification 
property of
this scheme is partially spoilt by the additional discrete symmetry required,
the non-renormalisable operators, and the extra Higgs singlet.

\section{mass spectrum}\label{cha:masses}
\subsection{$SU(2)_{\ell}$ unbroken}

The matter content and mass thresholds of a unified theory govern the running of the coupling constants.
It is thus important to elucidate at what energy scales the various particles gain mass and how 
they contribute to the renormalisation-group equations. 
We employ the same fermion and Higgs multiplet assignments as Ref.~\cite{BMW}, summarised
by Eqs.\ \ref{eqn:fermtransforms}-\ref{eqn:fermions} and \ref{eqn:higgstrans}.

Given that we are aiming for as natural a quartification model as possible, one needs to be aware of the 
most obvious approach in determining the Higgs VEV structure short of 
performing a minimisation analysis of a Higgs potential.
First, those Higgs fields of Eq.~\ref{eqn:higgstrans} which are not singlets under quark 
colour necessarily can not acquire VEVs, and we also naturally assume 
them to have mass of unification scale always. 
Thus the fields $\Phi_b$, $\Phi_d$, $\Phi_q$ and $\Phi_{q^c}$ have no influence on the renormalisation-group 
equations and can be ignored for now. For the remaining fields our policy is the following:
We first choose a symmetry breaking cascade. At a given 
stage in the symmetry breaking chain, those components that can acquire a VEV 
consistent with the symmetry breakdown pattern do so at that scale, 
and that the corresponding Higgs masses are also at that same scale. 

With this in mind, consider the breaking 
\begin{equation}\label{eqn:G4toSMSUl}
\centering
SU(3)_q \otimes SU(3)_L \otimes SU(3)_{\ell} \otimes SU(3)_R \stackrel{v}{\longrightarrow} 
SU(3)_q \otimes SU(2)_L \otimes SU(2)_{\ell} \otimes U(1)_Y \stackrel{u}{\longrightarrow}
SU(3)_q \otimes SU(2)_{\ell} \otimes U(1)_Q.
\end{equation}
The VEV pattern which achieves this is 
\begin{equation}\label{eqn:SUlVEV}
\langle \Phi_{\ell} \rangle = \left(\begin{array}{ccc}
0&0&u_{\ell}\\
0&0&0\\
0&0&v_{\ell}\end{array}\right)\qquad 
\langle \Phi_{\ell^c} \rangle = \left(\begin{array}{ccc}
0&0&0\\
0&0&0\\
v_{\ell^{c}_{1}}&0&v_{\ell^{c}_{2}}\end{array}\right)\qquad 
\langle \Phi_{a} \rangle = \langle \Phi_c^{\dagger} \rangle = 
\left(\begin{array}{ccc}u_{a1}&0&u_{a2}\\
0&u_{a3}&0\\
v_{a1}&0&v_{a2}\end{array}\right).
\end{equation}
(If we were to introduce intermediate steps in the breaking, then the $v$'s would be of 
different orders.) It is unfortunate that this VEV structure has
an intra-multiplet hierarchy, with entries of both order $u$ and $v$ contained 
in a single multiplet. 
We shall accept this as we think it would be more unnatural for the Higgs fields 
to get smaller VEVs than the 
symmetry breaking scheme requires. Solving this hierarchy problem would require additional Higgs fields (not necessarily 
in the same representations as those above), which would necessarily imply a larger Higgs potential
with a greater number of arbitrary parameters,
result in only a partially unified theory as in Ref.~\cite{Ray}, or 
require a completely different symmetry breaking mechanism, such as
the employment of inhomogeneous scalar field configurations~\cite{clash} 
or orbifold symmetry breaking in a brane-world setting~\cite{kawamura}(see Refs.~\cite{Alison}
and \cite{carone2} for applications of the former and latter, respectively, to trinification models). 

To give all Yukawa couplings even grounding, we remove the discrete symmetry employed in Ref.~\cite{BMW}, 
leaving four independent Yukawa interactions which can endow the fermions 
with mass. These are
\begin{eqnarray}\label{eqn:higgscouplings}
\centering
\lambda_q Tr[q^c \, q \, \Phi_{a} ] & 
\lambda_{\ell}  Tr [ \ell \, \ell^{c} \, \Phi_{c} ],\\
\lambda_{L} \epsilon^{jkl} \epsilon^{mnp}\ell^{jm} \ell^{kn} (\Phi_{\ell}^{\dagger})^{lp},&
\lambda_R \epsilon^{jkl} \epsilon^{mnp}(\ell^c)^{jm} (\ell^c)^{kn} (\Phi_{\ell^c})^{lp},
\end{eqnarray}
giving the quark mass term
\begin{equation}\label{eqn:quarkmass}
\mathcal{L}_{\rm quark\;mass}=
\lambda_q \left(\begin{array}{cc}d&h\end{array}\right) 
\left(\begin{array}{cc} u_{a 1}& u_{a 2}\\ v_{a 1} & v_{a 2}\end{array}\right) 
\left(\begin{array}{c} d^c \\ h^c \end{array}\right)
+ ( \lambda_q u_{a 3} ) u  u^c + H.c.
\end{equation}
The up quarks acquire electroweak scale Dirac masses, while the $d$ and $h$ quarks 
are mixed. Upon diagonalisation of this mass matrix, we have only one $Q = -1/3$ 
quark per family with electroweak scale mass to be identified as the down quark, and 
the exotic quark gains a GUT scale mass. Note that mixing between 
the $h$ and $d$ quarks is suppressed by $u/v$. 

The mass terms of the leptons are solely of Dirac nature. The $Q=-1$ charged leptons gain 
masses of electroweak order and do not mix with any other states. The liptons 
$x_1$, $x_2$, $y_1^c$, $y_2^c$, $z_1$ and $z_2$ have charge $Q=+1/2$ and pair up with 
the charge $-1/2$ liptons $x^c_1$, $x^c_2$, $y_1$, $y_2$, $z^c_1$ and $z^c_2$, 
to acquire GUT scale masses. The electrically neutral leptons $\nu$, $N$, $N^c$ and $\nu^c$ also 
only have Dirac mass terms. One sector is of GUT scale, identified as heavy neutral 
leptons, and the other, the ordinary neutrinos, is of electroweak scale. We see that we 
encounter the same problem as did Babu et al.\ with respect to obtaining a light neutrino mass. 

In summary, all SM particles including the Dirac neutrinos have electroweak scale masses, 
and all exotics have 
GUT scale masses. If the symmetry breaking occurs via intermediate scales, 
then the masses of the exotic particles will be at 
the unification or one of these intermediate scales. 

In determining the running of the gauge coupling constants, we must also 
know the full structure of the light Higgs spectrum at each stage of symmetry 
breaking. The VEV structure above neither provides enough information to define all the 
masses of the Higgs' components nor how many SM doublets there are. One is forced to make an 
assumption to deal with this, and, again, we adopt as natural a one as possible. 
The assumption chosen involves looking at the branchings, and 
particularly at what scale components branch away from those components that acquire VEVs. 
If a component gains a VEV, then the SM multiplet in which it is contained is taken to get a mass at the 
same scale. In the case where $SU(2)_{\ell}$ remains unbroken, there are SM multiplets which 
have no VEVs but are embedded within a quartification multiplet that does. We assume that 
these gain mass at the scale of the largest VEV in the quartification multiplet. For example, the VEV
\begin{equation} 
\langle \Phi_{\ell} \rangle = \left( \begin{array}{ccc}
0&0&u\\
0&0&0 \\
0&0&v \end{array} \right)
\end{equation}
implies that the components $(\Phi_{\ell})^1_3,(\Phi_{\ell})^2_3$ have masses of order 
$u$, while the remaining components all have mass of order $v$.
This gives us seven candidate light Higgs doublets at the SM level: one from $\Phi_\ell$
and three each from $\Phi_a$ and $\Phi_c$.  The Higgs doublet multiplicity has
a beneficial effect on the achievement of gauge coupling constant unification~\cite{Willenbrock}.    
Although the Higgs sector of our models has been burdened with these assumptions,
we have avoided the introduction of an additional discrete symmetry.

\subsection{$SU(2)_{\ell}$ broken}

The Higgs fields of Eq.~\ref{eqn:higgstrans} also have the capacity to break the 
leptonic colour symmetry completely, leaving no residual $SU(2)_\ell$ gauge group unbroken~\cite{lbroken}.
Consider the breaking cascade 
\begin{equation}\label{eqn:quarttosm}
\centering
SU(3)_q \otimes SU(3)_L \otimes SU(3)_{\ell} \otimes SU(3)_R \stackrel{v}{\longrightarrow} 
SU(3)_q \otimes SU(2)_L \otimes U(1)_Y \stackrel{u}{\longrightarrow}
SU(3)_q \otimes U(1)_Q.
\end{equation}
With this symmetry breaking pattern, the electric charge generator is given by
\begin{equation}\label{eqn:Q2}
\centering
Q=I_{3L}-\frac{Y_L}{2}+I_{3\ell}-\frac{Y_\ell}{2}+I_{3R}-\frac{Y_R}{2}.
\end{equation}
Notice that the three spontaneously broken $SU(3)$ factors contribute in a
symmetric manner to the electric charge generator.
This breakdown pattern is accomplished by Higgs fields obtaining VEVs of the form
\begin{equation}\label{eqn:VEV2}
\centering
\langle \Phi_{\ell} \rangle = \left(\begin{array}{ccc}
u_{\ell 1}&0&u_{\ell 2}\\
0&u_{\ell 3}&0\\
v_{\ell 1}&0&v_{\ell 2}\end{array}\right), \qquad 
\langle \Phi_{\ell^c} \rangle = 
\left(\begin{array}{ccc}
v_{\ell^c 1}&0&v_{\ell^c 2}\\
0&v_{\ell^c 3}&0\\
v_{\ell^c 4}&0&v_{\ell^c 5}\end{array}\right), \qquad 
\langle \Phi_{a} \rangle = \langle \Phi_{c}^{\dagger} \rangle = \left(\begin{array}{ccc}
u_{a1}&0&u_{a2}\\
0&u_{a3}&0\\
v_{a1}&0&v_{a2}\end{array}\right),
\end{equation}
where all Higgs components that can acquire a VEV at a given scale do so. 
The Higgs mass spectrum here is ``derived'' in a more obvious fashion than before. All members of SM 
multiplets  which get a VEV, acquire masses at that scale. 
This leaves nine light, left-handed Higgs doublets 
at the standard model level, three each from $\Phi_\ell$, $\Phi_a$ and $\Phi_c$. 

As before, Eq.~\ref{eqn:higgscouplings} describes the Yukawa couplings. 
Breaking leptonic colour completely has no impact on the quarks
as they are singlets under this gauge group: the quark masses are identical 
irrespective of whether or not $SU(2)_{\ell}$ is broken. 
The leptons, however, possess leptonic colour and their electric charges are altered 
due to the different electric charge generator of Eq.~\ref{eqn:Q2}, and their mass terms are greatly influenced by 
the different VEV pattern of Eq.~\ref{eqn:VEV2}.  The components that were previously
half-integrally charged liptons now acquire integral charges $Q = \pm 1,0$, so they
are no longer liptons but are instead charged and neutral heavy leptons.

The leptons with a charge of $+1$ are the $e^c$, $y_1^c$, $z_2$ and $x_2$
components of Eq.~\ref{eqn:fermions}. They mix and form Dirac mass terms 
with the charge $-1$ lepton components $e$, $y_1$, $z_2^c$ and $x_2^c$,
in the manner  
\begin{equation}\label{leptonmass2}
\left(\begin{array}{cccc} e& y_1 & z_2^c & x_2^c\end{array}\right)
\left(\begin{array}{cccc}
u_{a 3} & 0 & -u_{\ell 1} & v_{\ell 1} \\
0 & u_{a 3} & u_{\ell 2} & -v_{\ell 2} \\
-v_{\ell^c 1} & v_{\ell^c 4} & v_{a 2} & u_{a 2} \\
v_{\ell^c 2} & -v_{\ell^c 5} & v_{a 1} & u_{a 1}
\end{array}\right)
\left(\begin{array}{c}
e^c\\y_1^c\\z_2\\x_2
\end{array}\right) + H.c..
\end{equation}
There are three 
Dirac mass eigenvalues (per family) of GUT scale, and one eigenvalue (per family) 
of electroweak scale corresponding to the $e$, $\mu$ and $\tau$ masses. 

The leptonic components $N$, $N^c$, $\nu$, $\nu^c$, $x_1$, $x_1^c$, $y_2$, $y_2^c$, $z_1$ 
and $z_1^c$ are all neutral. Unlike the former case these ten leptons gain 
Majorana masses, as per 
\begin{equation}
\left(\begin{array}{cccccccccc}\!N&\!N^c&\!\nu&\!\nu^c&\!x_1&\!x_1^c&\!y_2&\!y_2^c&\!z_1&\!z_1^c\end{array}\!\right)
\left(\begin{array}{cccccccccc}
0  & v_{a2} & 0  & v_{a 1}& u_{L3} & 0 & u_{L 1} & 0 & 0 & 0\\
v_{a2} & 0  & u_{a 2} & 0 & 0 & v_{\ell^c 3} & 0 & v_{\ell^c 1} & 0 & 0 \\
0 & u_{a 2} & 0 &u_{a 1} & 0 & 0 &-u_{\ell 1} & 0 & -u_{L 3} & 0 \\
v_{a 1} & 0 & u_{a1} & 0 & 0 & 0 & 0& -v_{\ell^c 2} & 0 & -v_{\ell^c 3} \\
u_{\ell 3} & 0 & 0 & 0 & 0& u_{a 1}  &v_{\ell 2} & 0&0&0 \\
0 & v_{\ell^c 3} & 0 & 0 &u_{a 1} & 0 & 0 & v_{\ell^c 5} & 0 & 0 \\
u_{\ell 1} & 0 & -v_{\ell 1} & 0 & v_{\ell 2} & 0 & 0 & u_{a 3} & -u_{\ell
2} & 0 \\
0 & v_{\ell^c 1} & 0 & -v_{\ell^c 2} & 0 & v_{\ell^c 5} & u_{a 3} & 0 & 0
& 0 \\
0 & 0 & -u_{\ell 3} & 0 & 0 & 0 & -u_{\ell 2} & 0 & 0 & v_{a 2} \\
0 & 0 & 0 & -v_{\ell^c 3}& 0 & 0 & 0 & 0 &v_{a 2} & 0
\end{array}\!\right)\nonumber
\left(\begin{array}{c}N \\ N^c \\ \nu \\ \nu^c \\ x_1 \\ x_1^c \\ y_2 \\ y_2^c \\ z_1 \\ z_1^c\end{array}\right).
\label{eq:10x10}
\end{equation}
Nine of the resulting mass eigenvalues are of the order of the GUT scale, and the tenth 
has is a small mass of the order of $\frac{u^2}{v}$, which is precisely the 
mass scale that would result from a regular seesaw mechanism~\cite{seesaw}. 
This particle
displays the correct weak coupling with the electron to be identified as
the neutrino, and all interactions 
involving the light leptons with the heavy leptons are very suppressed. 
When intermediate scales are introduced, some of the order $v$ entries decrease in size
and thus some of the large eigenvalues also decrease.  One anticipates that this
raises the value of the smallest eigenvalue.

In summary, the VEV patterns of Eq.~\ref{eqn:VEV2} through the Yukawa coupling
terms provide large masses to exotic fermions, electroweak-scale masses for standard 
charged fermions, and a see-saw suppressed masses for the neutrinos.

\section{Renormalisation-group equations}

\subsection{$SU(2)_{\ell}$ unbroken}\label{cha:SU2l}

We begin by analysing the renormalisation-group equations for the schemes featuring
a remnant of the leptonic colour symmetry at low energy. 
There is no physical reason why the symmetry breaking has to directly proceed via 
Eq.~\ref{eqn:G4toSMSUl}. In fact in this case, without the restrictions 
proposed in Ref.~\cite{BMW} imposed on the Yukawa sector,
the gauge coupling constants only come within the vicinity of intersecting if 
the Higgs sector is enlarged significantly. 

An alternative to the one-step scenario 
is the introduction of intermediate symmetry breaking scales. 
There are four independent symmetry breaking routes from 
$G_4 \rightarrow SU(3)_q \otimes SU(2)_L \otimes SU(2)_{\ell} \otimes U(1)_Y$ 
which can be achieved with our Higgs sector. They are labelled as per:
\begin{eqnarray}
\centering
1. \: \: \: G_4 \: \: \: \stackrel{v}{\rightarrow} & SU(3)_q & \otimes SU(3)_L \otimes SU(2)_{\ell} \otimes SU(2)_R 
\otimes U(1)_{X_1} \\ \nonumber
\stackrel{w}{\rightarrow} & SU(3)_q & \otimes SU(2)_L \otimes SU(2)_{\ell} \otimes SU(2)_R 
\otimes U(1)_{X_2} \\ \nonumber
 \stackrel{x}{\rightarrow} & SU(3)_q & \otimes SU(2)_L \otimes SU(2)_{\ell}
\otimes U(1)_{Y} \label{eqn:cascade1a}\\ 
2. \: \: \: G_4 \: \: \: \stackrel{v}{\rightarrow} & SU(3)_q & \otimes SU(2)_L \otimes SU(2)_{\ell} \otimes SU(3)_R 
\otimes U(1)_{X_1} \\ \nonumber
\stackrel{w}{\rightarrow} & SU(3)_q & \otimes SU(2)_L \otimes SU(2)_{\ell} \otimes SU(2)_R 
\otimes U(1)_{X_2} \\ \nonumber
 \stackrel{x}{\rightarrow} & SU(3)_q & \otimes SU(2)_L \otimes SU(2)_{\ell}
\otimes U(1)_{Y}\\ 
3.\: \: \:  G_4 \: \: \: \stackrel{v}{\rightarrow} & SU(3)_q & \otimes SU(2)_L \otimes SU(3)_{\ell} \otimes SU(2)_R 
\otimes U(1)_{X_1} \\ \nonumber
\stackrel{w}{\rightarrow} & SU(3)_q & \otimes SU(2)_L \otimes SU(2)_{\ell} \otimes SU(2)_R 
\otimes U(1)_{X_2} \\ \nonumber
 \stackrel{x}{\rightarrow} & SU(3)_q & \otimes SU(2)_L \otimes SU(2)_{\ell}
\otimes U(1)_{Y}\\ 
4.\: \: \: G_4 \: \: \: \stackrel{v}{\rightarrow} & SU(3)_q & \otimes SU(2)_L \otimes SU(3)_{\ell} \otimes SU(2)_R 
\otimes U(1)_{X_1} \\ \nonumber
\stackrel{w}{\rightarrow} & SU(3)_q & \otimes SU(2)_L \otimes SU(3)_{\ell} 
\otimes U(1)_{X_2} \\ \nonumber
 \stackrel{x}{\rightarrow} & SU(3)_q & \otimes SU(2)_L \otimes SU(2)_{\ell}
\otimes U(1)_{Y}. \label{eqn:cascade4a}
\end{eqnarray}
For each cascade, the symmetry breaking is achieved by the VEVs of Eq.~\ref{eqn:SUlVEV} 
where the energy of the $v_i$ VEV entries are not uniform. Each of these cascades is of
course followed by electroweak symmetry breakdown.
As we now have several scales, the masses of the exotic fermions and Higgs bosons 
will stagger with energy and 
they will have varying contributions to the renormalisation-group equations. 
The exact nature of the VEV entries and at which scale the fermions gain masses are 
detailed in Table \ref{tab:masses}. Notice that the $x_1,x_2,y_1, y_2$ particles which had to 
be at the $TeV$ scale in Ref.~\cite{BMW}, can only potentially be light in the case of 
cascade four.
\begin{table}[h]
\begin{center}
\begin{tabular}{|c|c|c|c|c|}
\hline
Cascade & Energy scale of VEVs & Masses at $v$  & Masses at $w$  & Masses at $x$\\
\hline
1 & $ v_{R_2} \sim v,$ & $x_1^c, x_2^c, y_1^c,y_2^c$ & $h, h^c, z_1,z_1^c z_2, z_2^c,$ & none  \\
  & $v_{a2} \sim w, \; v_L \sim w,$ & &$N,N^c,x_1, x_2, y_1,y_2$ & \\
  & $v_{R_1} \sim x, \; v_{a1} \sim x$ & & & \\ 
\hline
2 & $v_L \sim v, $ & $x_1, x_2, y_1, y_2$ & $h, h^c, z_1,z_1^c z_2, z_2^c,$  & none \\
  & $v_{R_2} \sim w,\; v_{a2} \sim w, $ & &$N,N^c,x_1^c, x_2^c, y_1^c,y_2^c$ & \\
  & $ v_{R_1} \sim x, \; v_{a1} \sim x$ & & & \\
\hline
3 & $v_{a2} \sim v,$ & $h, h^c, z_1, z_2, N,$ & $x_1, x_2, y_1, y_2,$ & none \\
  & $v_L \sim w,\;  v_{R2} \sim w, $ & $z_1^c, z_2^c,N^c$ & $ x_1^c, x_2^c, y_1^c,y_2^c$ & \\
  & $v_{a1} \sim x, \; v_{R1} \sim x$ & & & \\
\hline
4 & $v_{a2} \sim v$,  & $h, h^c, z_1, z_2, N,$ & none & $x_1, x_2, y_1, y_2,$\\
  & $v_{a1} \sim w$  & $z_1^c, z_2^c,N^c$ & & $ x_1^c, x_2^c, y_1^c,y_2^c$ \\
  & $v_L \sim x, \; v_{R2} \sim x$ & & & \\
\hline
\end{tabular}
\end{center}
\caption{The energy scale of the VEV entries of Eq.~\ref{eqn:SUlVEV} 
with $v \geq w \geq x$ , and the enumeration of the 
fermion masses for the four cascades of Eqs.~\ref{eqn:cascade1a}-\ref{eqn:cascade4a} where 
the quartification gauge symmetry is broken down to $G_{SM} \otimes SU(2)_{\ell}$
in stages. Notice that 
cascade four is the only symmetry breaking route that will potentially permit light 
liptons.}
\label{tab:masses}
\end{table}

Defining the fine-structure constants as $\alpha_q$, $\alpha_L$, $\alpha_{\ell}$ and  
$\alpha_Y$, respectively, for quark colour, weak $SU(2)_L$, leptonic colour and hypercharge, the 
one-loop renormalisation-group equations which describe their evolution have the form
\begin{equation}
\centering
\frac{1}{\alpha_i (M_1)} = \frac{1}{\alpha_i (M_2)} + \frac{b_i}{2 \pi} \ln{ \left( \frac{M_1}{M_2} \right)}.
\end{equation}
$i=q,L,\ell,Y$, $M_{1,2}$ denote two mass scales of our theory, and the $b$ factor is given by 
\begin{equation}
\centering 
 b = - \frac{11}{3} T({\rm gauge \:\: bosons}) + \frac{2}{3} T ({\rm Weyl\:\: fermions}) 
+ \frac{1}{3} T({\rm complex \:\: scalars}).
\end{equation}
The $T$'s are group theoretical properties which depend on the gauge group representations
and are defined 
by the generators $\lambda^a$ in the representation $R$ as per 
\begin{equation}
\centering
Tr \left( \lambda^a \, \lambda^b \right) = T_R \delta^{ab}.
\end{equation}
At each stage of the symmetry breaking, $b$ harbours all knowledge of particles with masses 
lighter than that particular scale. Our labelling scheme for these factors is best illustrated
by an example: $b_{q_1}$ 
refers to the cumulative effect of fields which possess quark colour between $v$ and $w$; 
while $b_{q_2}$ is concerned with the energy range $x \leftrightarrow w$; and $b_{q_3}$ the range 
$M_{EW} \leftrightarrow x$.  The $b$-factors for the $\ell$, $L$ and $R$ sectors are denoted
similarly, with $q$ replaced by the appropriate subscript. 
The quantity $b_{u_j}, \, j=1,2,3$ shall denote the running of the $U(1)$ coupling and takes into account 
the normalisation of the generators defining $X_1, X_2$ and $Y$. The generator of hypercharge is 
taken to be the conventional embedding and is given by 
\begin{equation}
\centering
Y = I_{3R} + \frac{2}{\sqrt{3}} \left( \lambda_{L8} + \lambda_{\ell 8} + \lambda_{R8} \right),
\end{equation}
where the $\lambda$'s are the usual Gell-Mann generators.   

For all cascades, the running of the $\alpha_q , \alpha_L$ and $\alpha_{\ell}$ constants have the 
same generic form given by 
\begin{equation}
\centering
\frac{1}{\alpha_i (v)} = \frac{1}{\alpha_i (M_{EW})} - \frac{b_{i_1}}{2 \pi} \ln{ \left( \frac{v}{M_{EW}} \right)} 
+  \frac{b_{i_1}-b_{i_2}}{2 \pi} \ln{ \left( \frac{w}{M_{EW}} \right)}
+ \frac{b_{i_2}-b_{i_3}}{2 \pi} \ln{\left( \frac{x}{M_{EW}} \right)}.
\end{equation}
This arises as the $SU(2)_{L,\ell}$ groups have the same coupling constants
as $SU(3)_{L,\ell}$ because of 
the way in which these subgroups are embedded within their parent $SU(3)$. At each scale, the 
coupling constants are analysed and we can determine a relationship between the 
the different energy scales $v,w,x$ and the values of the fine-structure constants at 
the electroweak scale $M_{EW}$.
The evolutions of the $U(1)$ coupling constants are different for each cascade
as they depend on the specific linear combinations of generators defining $X_1$ and $X_2$, 
and so the forms of the renormalisation group equations
depend on the symmetry breaking pattern. The relationship between the
$U(1)$ electroweak-scale fine structure constant and
the electroweak-scale coupling constants for $SU(2)_\ell$ and $SU(2)_L$ is   
\begin{eqnarray}
\centering
\frac{1}{\alpha_Y} & = & \frac{1}{3 \alpha_{\ell}} + \frac{5}{3 \alpha_L} 
+ \frac{3 (b_{u_1} + b_{R_1})-b_{\ell_1}-4 b_{L_1}}{6 \pi} \ln{\left( \frac{v}{M_{EW}}\right)} \nonumber\\
& + & \frac{3 (b_{u_2} + b_{R_2}- b_{u_1} -  b_{R_1}) + b_{\ell_1}- b_{\ell_2} + 4 b_{L_1}-
5 b_{L_2}}{6 \pi} \ln{\left( \frac{w}{M_{EW}}\right)} \\ \nonumber
& + & \frac{3 ( b_{u_3} -  b_{u_2} - b_{R_2}) + b_{\ell_2}- b_{\ell_3}+ 5 ( b_{L_2}-
b_{L_3})}{6 \pi} \ln{\left(\frac{x}{M_{EW}}\right)}
\end{eqnarray}
for cascade one, but takes a different form for the other cascades.
The experimental values for the fine-structure constants at $M_{EW}$ are~\cite{PDG}
\begin{equation}
\centering
\alpha_q = 0.1172, \qquad \alpha_L = 0.0338, \qquad \alpha_Y = 0.0102,
\end{equation}
where remember we have absorbed the normalisation into the $b_{u_j}$'s.

In this section we shall focus on the renormalisation-group equation analysis for cascade one and present only a
 summary of results for the other symmetry breaking routes.
The full details of their equations are relegated to appendix \ref{cha:appendix1}. 
After the first stage of symmetry breaking, the particles $x_1^c$, $x_2^c$, $y_1^c$ and $y_2^c$ 
gain mass and the light Higgs spectrum is
\begin{eqnarray}
\centering
& \Phi_{\ell}& \sim \left( \mathbf{1,3,2,1} \right) \left( \frac{1}{3} \right) \oplus
\left( \mathbf{1,3,1,1} \right) \left(-\frac{2}{3}\right), \qquad \qquad \qquad
\Phi_{\ell^c} \sim \left( \mathbf{1,1,1,2} \right) (1), \\ 
& \Phi_a & \sim  \left( \mathbf{1,3,1,2} \right) \left(\frac{1}{3}\right) \oplus
\left( \mathbf{1,3,1,1} \right) \left(-\frac{2}{3}\right), \qquad \Phi_c \sim \Phi_a^{\dagger}.
\end{eqnarray} 
The remaining exotic fermions gain masses at the $w$ scale and the 
light Higgs spectrum is 
\begin{eqnarray}
\centering
& \Phi_{\ell}& \sim \left( \mathbf{1,2,1,1} \right) (-1), \qquad
\Phi_{\ell^c} \sim \left( \mathbf{1,1,1,2} \right) (1), \\ 
& \Phi_a & \sim  \left( \mathbf{1,2,1,2} \right) (0) \oplus
\left( \mathbf{1,2,1,1} \right) (-1) \oplus
\left( \mathbf{1,1,1,2} \right) (1), \qquad \Phi_c \sim \Phi_a^{\dagger}.
\end{eqnarray} 
After the final breaking particles of the minimal SM and $\nu^c$ are massless and 
\begin{eqnarray}
\centering
& \Phi_{\ell}& \sim \left( \mathbf{1,2,1} \right) (-1), \\
& \Phi_a & \sim  \left( \mathbf{1,2,1} \right) (-1) \oplus
\left( \mathbf{1,2,1} \right) (1) \oplus
\left( \mathbf{1,2,1} \right) (-1), \qquad \Phi_c \sim \Phi_a^{\dagger},
\end{eqnarray}
giving us seven SM Higgs doublets. 

Defining $N_H$ to be the multiplicity of the Higgs fields in 
Eq.~\ref{eqn:higgstrans}\footnote{We take the same multiplicity $N_H$ for 
{\it{both}} of the $\mathbf{36}$'s representing our Higgs fields. This
can obviously be generalised, but we shall have no need to do this
because we shall focus only on the simplest possible 
schemes in which the multiplicity of each representation is precisely $N_H=1$.
Happily, it turns out $N_H > 1$ is not required.}
and summing over three generations of fermions, this spectrum of particles 
defines the values of the $b$ quantities as
\begin{eqnarray}
\begin{array}{ccccc}
b_{q_1}  =  -5,&  b_{L_1} = - 5 + \frac{3 N_H}{2},& 
b_{{\ell}_1} = - \frac{10}{3} + \frac{N_H}{2}, &
b_{R_1} = -\frac{10}{3} + \frac{7 N_H}{6}, & b_{u_1} = 4 + \frac{2 N_H}{3},\\ 
b_{q_2} =  -7, & b_{L_2} = - \frac{10}{3} + \frac{7 N_H}{6}, &
b_{{\ell}_2} = -\frac{22}{3},&  b_{R_2}= -\frac{10}{3}+\frac{7 N_H}{6}, &
b_{u_2} = \frac{8}{3} + N_H, \\ 
b_{q_3} =  b_{q_2}, & b_{L_3}=b_{L_2}, & b_{{\ell}_3} = b_{{\ell}_2}, & &
b_{u_3}=\frac{20}{3}+\frac{7 N_H}{6}.\end{array}
\end{eqnarray}

Substituting these numbers into the renormalisation-group equations, we have
\begin{eqnarray}
\centering
\frac{1}{\alpha_q (v)} & =& \frac{1}{\alpha_q} + \frac{5}{2 \pi} \ln{ \left( \frac{v}{M_{EW}} \right) } 
+ \frac{1}{\pi} \ln{ \left( \frac{w}{M_{EW}} \right) }, \\ 
\frac{1}{\alpha_L (v)} & = &\frac{1}{\alpha_L} + \frac{10-3 N_H}{4 \pi} \ln{ \left( \frac{v}{M_{EW}} \right) } 
+ \frac{N_H - 5 }{6 \pi} \ln{ \left( \frac{w}{M_{EW}} \right) }, \\ 
\frac{1}{\alpha_{\ell} (v)} & =& \frac{1}{\alpha_{\ell}} + \frac{20-3 N_H}{12 \pi}
\ln{ \left( \frac{v}{M_{EW}} \right) } 
+ \frac{8 + N_H }{4 \pi} \ln{ \left( \frac{w}{M_{EW}} \right) }, \\ 
\frac{1}{\alpha_Y} & = & \frac{1}{3 \alpha_{\ell}} + \frac{5}{3 \alpha_L} 
+ \frac{76-3 N_H}{18 \pi} \ln{ \left( \frac{v}{M_{EW}} \right) }
+ \frac{5 N_H-10 }{18 \pi} \ln{ \left( \frac{w}{M_{EW}} \right) }
+ \frac{22-3 N_H }{6 \pi} \ln{ \left( \frac{x}{M_{EW}} \right) }.
\end{eqnarray}
With these inputs, we have a large degree of freedom in the unification of the couplings, with 
the simplest scheme being that in which $N_H=1$. 

Given that we have no particles gaining mass at the final stage of breaking, the scale $x$ 
affects only the evolution of the 
hypercharge fine-structure constant.
As a result, this scale can be as low as a few $TeV$ and as large as $6 \times 10^7\ GeV$ 
without spoiling the unification. Taking 
$x \equiv x_{min} \sim 1\ TeV$, unification occurs for 
\begin{equation}
\centering
w \sim 2.7 \times 10^{12}\ GeV, \qquad v \sim 1.2 \times 10^{17}\ GeV.
\end{equation}
At this unification scale, the value of the fine-structure constant for our unified theory is 
$\alpha_{G_4}^{-1} \sim 43.85$, giving $\alpha_{\ell} \sim 0.0912$ at the electroweak scale, 
which has a value between the weak $SU(2)_L$ and the strong couplings.
If we allow $x$ to increase 
in energy, then the GUT scale decreases and the $w$ scale increases until $v = w \sim 7.5 \times 10^{13}\ GeV$ 
at $x_{max} \sim 6.4 \times 10^7\ GeV$. This unification scheme gives $\alpha_{G_4}^{-1} \sim 39.02$ 
and $\alpha_{\ell} \sim 0.1408$.
It is interesting to note here that the scale $w$ cannot be low for unification purposes, and 
subsequently all our exotic fermions will be heavy, leaving only the $SU(2)_{\ell}$ gauge bosons 
and additional Higgs fields as light particles foreign to the standard model. 

Unification of the gauge coupling constants can also be obtained for the other three breaking patterns 
with the range of possible, consistent scales summarised in Table \ref{tab:su2unbroken}. 
\begin{table}[h]
\begin{center}
\begin{tabular}{|c|c|c|c|c|c|}
\hline
Cascade & $x$ & $w$ & $v$ & $\alpha_{G_4}^{-1}$ & $\alpha_{\ell}$ \\
\hline
1  & $x_{min}\sim 1 TeV$ & $2.7 \times 10^{12} GeV$ & $ 1.2 \times 10^{17} GeV$ & 43.85 & 0.0912 \\
  & $x_{max}\sim 6.4 \times 10^7 GeV$ & $7.5 \times 10^{13} GeV$ & $ 7.5 \times 10^{13} GeV$ & 39.02 & 0.1408 \\
\hline
2  & $x_{min} \sim 6.5 \times 10^5 GeV$ & $6.5 \times 10^{5} GeV$ & $3.9 \times 10^{19} GeV$ & 43.55 & 0.0526 \\
   & $x_{max} \sim 6.5 \times 10^{7} GeV$ & $7.4 \times 10^{13} GeV$ & $7.4 \times 10^{13} GeV$ & 39.01  & 0.1407 \\
 \hline
3  & $x_{min} \sim 6.3 \times 10^7 GeV$ & $ 7.7 \times 10^{13} GeV$ & $ 7.7 \times 10^{13} GeV$ & 39.02 & 0.1412 \\
   & $x_{max} \sim 4.9 \times 10^{10} GeV$ & $ 4.9 \times 10^{10} GeV$ & $ 7 \times 10^{12} GeV$ & 36.35  & 0.1210 \\
\hline
4  & $x_{min} \sim 6.2 \times 10^8 GeV$ & $1.7 \times 10^{12} GeV$ & $1.7 \times 10^{12} GeV $ & 34.77 & 0.111 \\
   & $x_{max} \sim 4.8 \times 10^{10} GeV$ & $4.8 \times 10^{10} GeV$ & $7 \times 10^{12} GeV$ & 36.35  & 0.1208 \\
\hline
\end{tabular}
\end{center}
\caption{Range of energy scales of symmetry breaking that yield unification of the gauge coupling constants. 
There is only one scenerio which allows for a $TeV$ level breaking scale, while the scales 
offered by the other choices are quite similar to each other.}
\label{tab:su2unbroken}
\end{table}
Unlike the first case, all the intermediate symmetry breaking scales for the other patterns must be high. 
The lowest the final intermediate breaking scale can be is
about $6 \times 10^5\ GeV$ in option two, which is significantly 
higher than the electroweak scale. Furthermore, this choice requires the unification scale 
to be larger than the Planck scale, which is unacceptable. So, realistically, 
in this scheme we would have to consider higher values of $x$ so as to lower the unification scale.
As a result, the theories prescribed by cascades two, three and four will 
contain very heavy exotic particles that do not lie within reach of future colliders.
Note also that the value of the fine-structure constant for leptonic colour $SU(2)_\ell$
at the electroweak scale is generally always 
larger than that describing quark colour.

Recall that the fourth symmetry breaking option is the only route that
returns exotic particles with masses of order $x$. However, the value of $x$ can 
only be pushed down in energy to $6 \times 10^8\ GeV$ if unification is to be preserved. 
Consequently, the possible existence of new low-energy phenomenology suggested by the presence of exotic 
fermions at this last stage of breaking is denied by the demands placed on the energy scale 
by the unification of the gauge coupling constants. 

We shall comment further on the phenomenology of our models in Sec.~\ref{cha:phenomenology}

\subsection{$SU(2)_{\ell}$ broken} \label{cha:RGE-SU(2)broken}

By demanding that the leptonic colour symmetry is broken entirely, the number of symmetry 
breaking routes from the quartification gauge group to the SM broadens to eight independent cascades. 
These are labelled as per:
\begin{eqnarray}
\centering
1. \;\; G_4 & \stackrel{v}{\rightarrow} & 
SU(3)_q \otimes SU(3)_L \otimes SU(2)_{\ell} \otimes SU(2)_R \otimes U(1)_{X_1} \nonumber \\
& \stackrel{w}{\rightarrow} & SU(3)_q \otimes SU(2)_L \otimes SU(2)_{\ell} \otimes SU(2)_R \otimes U(1)_{X_2} \nonumber \\
& \stackrel{x}{\rightarrow} & SU(3)_q \otimes SU(2)_L \otimes SU(2)_{\ell} \otimes U(1)_{X_3} \nonumber \\
& \stackrel{y}{\rightarrow} & SU(3)_q \otimes SU(2)_L \otimes U(1)_{Y} \\
2. \;\; G_4 & \stackrel{v}{\rightarrow} & 
SU(3)_q \otimes SU(3)_L \otimes SU(2)_{\ell} \otimes SU(2)_R \otimes U(1)_{X_1} \nonumber \\
& \stackrel{w}{\rightarrow} & SU(3)_q \otimes SU(2)_L \otimes SU(2)_{\ell} \otimes SU(2)_R \otimes U(1)_{X_2} \nonumber \\
& \stackrel{x}{\rightarrow} & SU(3)_q \otimes SU(2)_L \otimes SU(2)_R \otimes U(1)_{X_3} \nonumber \\
& \stackrel{y}{\rightarrow} & SU(3)_q \otimes SU(2)_L \otimes U(1)_{Y} \\
3. \;\; G_4 & \stackrel{v}{\rightarrow} & 
SU(3)_q \otimes SU(3)_L \otimes SU(2)_{\ell} \otimes SU(2)_R \otimes U(1)_{X_1} \nonumber \\
& \stackrel{w}{\rightarrow} & SU(3)_q \otimes SU(3)_L \otimes U(1)_{X_2} \nonumber \\
& \stackrel{y}{\rightarrow} & SU(3)_q \otimes SU(2)_L \otimes U(1)_{Y} \\
4. \;\; G_4 & \stackrel{v}{\rightarrow} & 
SU(3)_q \otimes SU(2)_L \otimes SU(3)_{\ell} \otimes SU(2)_R \otimes U(1)_{X_1} \nonumber \\
& \stackrel{w}{\rightarrow} & SU(3)_q \otimes SU(2)_L \otimes SU(2)_{\ell} \otimes SU(2)_R \otimes U(1)_{X_2} \nonumber \\
& \stackrel{x}{\rightarrow} & SU(3)_q \otimes SU(2)_L \otimes SU(2)_{\ell} \otimes U(1)_{X_3} \nonumber \\
& \stackrel{y}{\rightarrow} & SU(3)_q \otimes SU(2)_L \otimes U(1)_{Y} \\
5. \;\; G_4 & \stackrel{v}{\rightarrow} & 
SU(3)_q \otimes SU(2)_L \otimes SU(3)_{\ell} \otimes SU(2)_R \otimes U(1)_{X_1} \nonumber \\
& \stackrel{w}{\rightarrow} & SU(3)_q \otimes SU(2)_L \otimes SU(2)_{\ell} \otimes SU(2)_R \otimes U(1)_{X_2} \nonumber \\
& \stackrel{x}{\rightarrow} & SU(3)_q \otimes SU(2)_L \otimes SU(2)_R \otimes U(1)_{X_3} \nonumber \\
& \stackrel{y}{\rightarrow} & SU(3)_q \otimes SU(2)_L \otimes U(1)_{Y} \\
6. \;\; G_4 & \stackrel{v}{\rightarrow} & 
SU(3)_q \otimes SU(2)_L \otimes SU(3)_{\ell} \otimes SU(2)_R \otimes U(1)_{X_1} \nonumber \\
& \stackrel{w}{\rightarrow} & SU(3)_q \otimes SU(2)_L \otimes SU(3)_{\ell} \otimes U(1)_{X_2} \nonumber \\
& \stackrel{x}{\rightarrow} & SU(3)_q \otimes SU(2)_L \otimes SU(2)_{\ell} \otimes U(1)_{X_3} \nonumber \\
& \stackrel{y}{\rightarrow} & SU(3)_q \otimes SU(2)_L \otimes U(1)_{Y} \\
7. \;\; G_4 & \stackrel{v}{\rightarrow} & 
SU(3)_q \otimes SU(2)_L \otimes SU(2)_{\ell} \otimes SU(3)_R \otimes U(1)_{X_1} \nonumber \\
& \stackrel{w}{\rightarrow} & SU(3)_q \otimes SU(2)_L \otimes SU(2)_{\ell} \otimes SU(2)_R \otimes U(1)_{X_2} \nonumber \\
& \stackrel{x}{\rightarrow} & SU(3)_q \otimes SU(2)_L \otimes SU(2)_{\ell} \otimes U(1)_{X_3} \nonumber \\
& \stackrel{y}{\rightarrow} & SU(3)_q \otimes SU(2)_L \otimes U(1)_{Y} \\
8. \;\; G_4 & \stackrel{v}{\rightarrow} & 
SU(3)_q \otimes SU(2)_L \otimes SU(2)_{\ell} \otimes SU(3)_R \otimes U(1)_{X_1} \nonumber \\
& \stackrel{w}{\rightarrow} & SU(3)_q \otimes SU(2)_L \otimes SU(2)_{\ell} \otimes SU(2)_R \otimes U(1)_{X_2} \nonumber \\
& \stackrel{x}{\rightarrow} & SU(3)_q \otimes SU(2)_L \otimes SU(2)_R \otimes U(1)_{X_3} \nonumber \\
& \stackrel{y}{\rightarrow} & SU(3)_q \otimes SU(2)_L \otimes U(1)_{Y},
\end{eqnarray}
where the generator of hypercharge now has the form 
\begin{equation}
\centering
Y=I_{3R}+I_{3 \ell} +\frac{2}{\sqrt{3}} \left( \lambda_{L8} + \lambda_{\ell 8} +\lambda_{R8} \right).
\end{equation}
Of these eight choices, there are seven which can deliver unification 
of the gauge coupling constants.
Cascade three does not offer a viable model assuming a minimal Higgs sector is used, so we eliminate it from
further consideration.
In all models, the light 
particle spectrum consists of the standard model 
particles and nine candidate SM Higgs doublets. 
The exotic fermions gain masses either of order $v$ or $w$ in all cases except for cascade six. 
The VEV structure of cascade six endows the particles 
$x_1$,  $x_2$,  $y_1$,  $y_2$,  $x_1^c$,  $x_2^c$,  $y_1^c$ and $y_2^c$ with masses at the $x$ scale, 
which at first sight could potentially result in lighter masses than the other scenarios.

For all cascades, the full details of the particle spectra, including the Higgs VEV patterns 
instigating the breaking, 
and the analysis of the renormalisation-group equations, are 
contained in appendix~\ref{cha:appendix2}. Table~\ref{tab:unifscales} provides 
a summary of results for the possible ranges of 
energy scales which give unification of the gauge coupling constants for $N_H=1$. 

Only a subset of these seven symmetry breaking schemes allow for a 
flexible range of unification and intermediate breaking scales. 
The final breaking to the SM gauge group can be as low as a $TeV$ for six of these seven options, 
with cascades one, two, seven and eight demanding that this scale be precisely of $TeV$ order. 
In fact, $y \equiv y_{max} \sim 7.1 \times 10^2\ GeV$ is the highest scale at which this breaking 
occurs for these options. This choice of $y$-scale offers only two intermediate scales with unification 
requiring $x=y$ and $w=v \sim 1.3 \times 10^{13}\ GeV$. The symmetry breaking patterns
for these four cascades thus 
become equivalent, reducing to 
$G_4 \rightarrow SU(3)_q \otimes SU(2)_L \otimes SU(2)_{\ell} \otimes SU(2)_R \otimes U(1) 
\rightarrow SU(3)_q \otimes SU(2)_L \otimes U(1)_Y$.

Unification can still be achieved in options four and five if the $y$ scale is as high as $\sim 10^6\ GeV$. 
Furthermore, once at the $SU(3)_q \otimes SU(2)_L \otimes SU(2)_{\ell} \otimes SU(2)_R \otimes U(1)$ 
level, the choice of which $SU(2)_{\ell,R}$ factor to break first has no significant influence on 
the outcome of the unification and intermediate scales.
It turns out that for all viable $y$ values, $x$ must be very close to $y$. 
When the final breaking occurs at $y_{min} \sim 1\ TeV$, then 
the unification scale is of order $10^{13}\ GeV$, whereas at $y_{max} \sim 10^6\ GeV$, the unification 
scale is at a lower energy, of order $10^{11}\ GeV$, which could potentially be more dangerous with 
respect to proton decay. Nevertheless, for all these cascades, the highest the unification 
scale can be is $\sim 10^{13}\ GeV$ which is much lower than the GUT energies possible when 
$SU(2)_{\ell}$ remains unbroken at low energy. 

Cascade six affords the most tantalising spectrum of masses for the exotic fermions, 
with masses of order $x$ resulting from the breaking 
$G_{SM} \otimes SU(3)_{\ell} \rightarrow G_{SM} \otimes SU(2)_{\ell}$. 
This scale can be as low as $\sim 10^4\ GeV$ with unification 
occurring for  
\begin{equation}
\centering
y_{min} \sim 1\ TeV, \qquad x \sim 8.8 \times 10^3\ GeV, \qquad w=v\equiv v_{min} \sim 3.6 \times 10^{10}\ GeV.
\end{equation}
A $10\ TeV$ scale for some of the exotic fermion masses provides hope for possible discovery
at the LHC.  However, this choice also requires the unification
scale $v \sim 10^{10}\ GeV$, which may be low enough to be troubling with respect to 
proton decay. If we allow $x$ to increase, then we obtain
\begin{equation}
\centering
y_{min} \sim 1\ TeV, \qquad x = w \sim 4.2 \times 10^7\ GeV, \qquad v\equiv v_{max} \sim 3.8 \times 10^{11}\ GeV.
\end{equation}
There is flexibility in $y$; it can be pushed up to 
\begin{equation}
y_{max} =x=w \sim 1.2 \times 10^6 GeV, \qquad v\sim 1.4 \times 10^{11}GeV, \qquad \alpha^{-1}_{G_4} = 32.02.
\end{equation}
With this choice, it becomes equivalent to the upper bound of unification for cascades four and five, 
with the symmetry breaking now described by 
$G_4 \rightarrow SU(3)_q \otimes SU(2)_L \otimes SU(3)_{\ell} \otimes SU(2)_R \otimes U(1)_{X_1} \rightarrow G_{SM}$.  
Since the $x$-scale is now at about $10^6\ GeV$, we see that while this cascade is consistent
with exotic fermion masses of about $10\ TeV$, they can also be significantly higher without
spoiling unification.

\begin{table}
\begin{center}
\begin{tabular}{|c|c|c|c|c|c|c|}
\hline
 & $y$ & $x$ & $w$ & $v$ & $\alpha_{G_4}^{-1}$ \\
\hline
1 and 2 &  $y_{max} \sim 7.1 \times 10^2 GeV$ & $7.1 \times 10^2 GeV$ &
$1.3 \times 10^{13} GeV$  & $1.3 \times 10^{13} GeV$& 37.05\\
\hline
4 and 5 & $y_{min} \sim 1 TeV$ & $1TeV$ & $6.2 \times 10^{12} GeV$ &$v_{max} \sim 1.1 \times 10^{13} GeV$ & 36.82 \\
  &    & $4.2 \times 10^7 GeV$ & $4.2 \times 10^7 GeV$ &$v_{max} \sim 3.8 \times 10^{11} GeV$ & 33.11\\
& $y_{max} \sim 1.2 \times 10^6 GeV$ & $1.2 \times 10^6 GeV$  &$1.2 \times 10^6 GeV$ &
$1.4 \times 10^{11} GeV$ & 32.02 \\
\hline
6 & $y_{min} \sim 1 TeV $ & $8.8 \times 10^{3} GeV$ &$3.6 \times 10^{10} GeV$ &
$v_{min} \sim 3.6 \times 10^{10} GeV$ & 30.48 \\
  &    & $4.2 \times 10^7 GeV$ & $4.2 \times 10^7 GeV$ &$v_{max} \sim 3.8 \times 10^{11} GeV$ & 33.11\\
 & $y_{max} \sim 1.2 \times 10^{6} GeV$ & $1.2 \times 10^{6} GeV$ &$1.2 \times 10^{6} GeV$ &
$1.4 \times 10^{11} GeV$ & 32.02 \\
\hline
7 and 8 & $y_{max} \sim 7.1 \times 10^2 GeV$&$ 7.1 \times 10^2 GeV$ & $1.3 \times 10^{13} GeV$ &
$1.3 \times 10^{13} GeV$ & 37.05\\
\hline
\end{tabular}
\end{center}
\caption{The range of energies for the symmetry breaking scales that will 
consistently give unification of the gauge coupling constants, when $N_H=1$. 
Four of the schemes become equivalent if unification is to be demanded, and the last 
stage of symmetry breaking has to occur below a $TeV$. The other three choices allow 
for a range in the intermediate scales while still preserving unification. 
When $y_{max}$ chosen for cascades four, five and six, they also become equivalent. }\label{tab:unifscales}
\end{table}

\section{phenomenology}\label{cha:phenomenology}

We now round out our discussion of the phenomenological consequences of the various schemes above.
It is beyond the scope of this paper to provide a rigorous quantitative analysis of
phenomenological bounds for all of these models, so our remarks shall be qualitative and
our analysis necessarily incomplete.

We first deal with the four cascades featuring a remnant $SU(2)_\ell$.  All of them feature
seven electroweak Higgs doublets.
We emphasise that this multiplicity is not due to duplication of the fundamental Higgs multiplets.
Rather, they are an integral part of the minimal Higgs sector required for quartification.
Obviously, questions about Higgs-induced flavour-changing neutral processes arise.  Without a
detailed analysis we can only make the simple remark that some of the doublets will have to
acquire $TeV$ scale masses or have somewhat small Yukawa coupling constants.  It is 
certainly interesting, though, that multiple Higgs doublets are a generic prediction of
quartification models.

Cascade one is the only
one that allows an intermediate scale as low as a $TeV$.  This scale is a right-handed weak-isospin
breaking scale, so the immediate phenomenological consequences are right-handed $W$ bosons
and a corresponding $Z'$ at the $TeV$ level.  Since this scale can be raised above $10^4\ TeV$
without spoiling unification, it is clear that it can be made phenomenologically acceptable.
However, it has no necessary new physics at LHC energies, apart from the
multi-Higgs doublet feature it shares with the other quartification schemes.  
Recall from the earlier discussion that it also has no $TeV$-scale exotic fermions.
The unification scale is in the range $10^{14-17}\ GeV$, so we would guess that it
is safe from too-rapid Higgs-induced proton decay.

Cascades two, three and four all have high $x$-scales, so their only characteristic $TeV$-level
feature is the seven Higgs doublets.  The unification scales lie in the range
$7 \times 10^{12-13}\ GeV$, which should be safe from a proton-decay point of view.

We now turn to the schemes having no leptonic colour remnant symmetry.  As noted
earlier, cascade three is unsuccessful and hence discarded.  All the cascades feature
nine electroweak Higgs doublets.

The requirement of
unification makes cascades one, two, seven and eight identical, with the lowest
breaking scale being at about $700\ GeV$.  This scheme is possibly ruled out,
because it results in quite light right-handed $W$-bosons and other light gauge
particles including $Z'$ states.  The $700\ GeV$ scale follows from the central values for
the electroweak-scale gauge coupling constants; it can be pushed up to the $TeV$ range
by varying these values within the experimental error range.
All the exotic fermions, however, are quite heavy,
gaining masses at the $w$-scale which is about $10^{13}\ GeV$.  If detailed
study were to show it is not yet falsified, then it would be an interesting 
situation in regards to possible discovery of new gauge bosons below $1\ TeV$.
The unification scale of $10^{13}\ GeV$ may be sufficient to suppress proton decay.

Cascades four and five can feature, respectively, $SU(2)_\ell$ or $SU(2)_R$ gauge bosons at
the $TeV$-level, although they need not.  It would be interesting to study
their GUT-scale proton-decay phenomenology, as a lower $y$-value implies a higher
GUT-scale, as summarised in Table~\ref{tab:unifscales}.  Conceivably, the suppression 
of Higgs-induced proton decay might
favour new $TeV$-scale physics for these cascades.

As already noted, cascade six is unique in that it can have new fermions
at the relatively low scale of about $10\ TeV$. If so, this would be
correlated with a low breaking scale for $SU(2)_\ell$ and hence the presence
of exotic gauge bosons coupling leptons to exotic leptons.  There is also
extended neutral current phenomenology.  The danger for cascade six is the
low range for the unification scale, which is not allowed to be much
higher than $10^{11}\ GeV$.  

In summary, there is obviously a wealth of phenomenology to be
explored within these schemes, both at the $TeV$-scale and at the GUT-scale.
The overall impression is that the schemes that totally break leptonic colour
are more constrained, either from $TeV$-scale considerations or from the GUT
regime or both.  This makes them more exciting, more easily tested; some are
possibly already ruled out.  Note that we have not yet 
attempted a systematic study of the Higgs-induced proton decay question,
so our concerns about some of the lower unification scales are generic rather 
than specific.

We have also not yet attempted a study of the Higgs potential and the minimisation
conditions.  It almost goes without saying that all proposed quartification
schemes suffer from the gauge hierarchy problem.  In our opinion, however, the
overall framework has considerable appeal, despite this
standard defect common to all non-supersymmetric GUTs.

Another interesting topic for future work is neutrino mass generation for the
totally-broken leptonic colour scenarios, to understand the effect of the intermediate
scales on the see-saw suppression given by Eq.~\ref{eq:10x10}.

\section{conclusion}\label{cha:conclusion}

Quartification schemes offer an alternative route to grand unification.  They
are conceptually rather appealing, with the fundamental fermion and Higgs
multiplets taking relatively simple and elegant forms.  As we have shown
in this paper, there are a variety of symmetry breaking cascades consistent
with successful gauge coupling constant unification.  None of them require
supersymmetry, though all of them of course require intermediate scales.
The non-trivial result is that appropriate intermediate scales are a natural
possibility.  Our results add to the important observation of Babu, Ma and
Willenbrock \cite{BMW} that complete unification is possible in quartification models,
rather than having to settle for the partial unification originally proposed
by Joshi and Volkas \cite{Ray}.

The various schemes have different phenomenological consequences, though
all have the existence of several electroweak Higgs doublets as a feature.
This multiplicity is not due to a replication of fundamental Higgs multiplets,
but is rather an inherent feature of the minimal Higgs sector required
for quartification. Depending on the scheme, rich phenomenology at LHC energies
such as additional gauge bosons and fermions is possible and in some cases required.
In addition, the models may have Higgs-induced proton decay,
though detailed analyses of this and the new physics at the $TeV$-scale
have yet to be carried out.

\acknowledgments{This work was supported by the Australian Research Council,
the Commonwealth of Australia, and the University of Melbourne.}

\appendix
\section{RGEs for $SU(2)_{\ell}$ unbroken}\label{cha:appendix1}
Here we summarise the analysis of the renormalisation-group equations analysis for cascades two, three and four
within the class that leaves the leptonic colour remnant symmetry $SU(2)_\ell$ unbroken.

\subsection{Cascade 2}

The renormalisation-group equations are
\begin{eqnarray}
\centering
\frac{1}{\alpha_i (v)} &=& \frac{1}{\alpha_i (M_{EW})} - \frac{b_{i_1}}{2 \pi} \ln{ \left( \frac{v}{M_{EW}} \right)} 
+  \frac{b_{i_1}-b_{i_2}}{2 \pi} \ln{ \left( \frac{w}{M_{EW}} \right)}
+ \frac{b_{i_2}-b_{i_3}}{2 \pi} \ln{\left( \frac{x}{M_{EW}} \right)}, \qquad i=q,L,\ell \label{eqn:RGEgeneric} \\
\frac{1}{\alpha_Y} & = &
\frac{1}{3 \alpha_{\ell}} + \frac{5}{3 \alpha_L} 
+ \frac{3 b_{u_1} + 4 b_{R_1}-b_{\ell_1}-5 b_{L_1}}{6 \pi} \ln{ \left(\frac{v}{M_{EW}} \right)} \\ \nonumber
&+& \frac{3 b_{u_2} + 3 b_{R_2}-3 b_{u_1} - 4 b_{R_1} + b_{\ell_1}- b_{\ell_2} + 5 b_{L_1}-
5 b_{L_2}}{6 \pi} \ln{ \left( \frac{w}{M_{EW}}\right)} \\ \nonumber
& + &\frac{3 b_{u_3} - 3 b_{u_2} - 3 b_{R_2} + b_{\ell_2}- b_{\ell_3}+ 5 b_{L_2}-
5 b_{L_3}}{6 \pi} \ln{\left(\frac{x}{M_{EW}}\right)}.
\end{eqnarray}
As illustrated in Table \ref{tab:masses}, the leptons $x_1$, $x_2$, $y_1$ and $y_2$ have masses of order 
$v$ while the remaining exotic fermions gain mass at $w$. The {\it{light}} Higgs sector has the structure 
\begin{eqnarray}
\Phi_{\ell} & \stackrel{v}{\rightarrow} & \left( \mathbf{ 1,2,1,1 } \right) (-1) 
\stackrel{x}{\rightarrow} \left( \mathbf{ 1,2,1 } \right) (-1), \\ \nonumber
\Phi_{\ell^c} & \stackrel{v}{\rightarrow} & \left( \mathbf{ 1,1,2,\overline{3} } \right) \left(-\frac{1}{3}\right) 
\oplus  \left( \mathbf{ 1,1,1,\overline{3} } \right) \left(\frac{2}{3}\right)
\stackrel{w}{\rightarrow} \left( \mathbf{ 1,1,1,2 } \right) (1), \stackrel{x}{\rightarrow} nothing, \\ \nonumber  
\Phi_a & \stackrel{v}{\rightarrow} & \left( \mathbf{ 1,2,1, \overline{3} } \right) \left(-\frac{1}{3}\right) 
\oplus  \left( \mathbf{ 1,1,1,\overline{3} } \right) \left(\frac{2}{3}\right)
\stackrel{w}{\rightarrow} \left( \mathbf{ 1,2,1,2 } \right) (0) \oplus \left( \mathbf{ 1,1,1,2 } \right) (1)
\oplus \left( \mathbf{ 1,2,1,1 } \right) (-1), \\ \nonumber  
& \stackrel{x}{\rightarrow} & \left( \mathbf{ 1,2,1 } \right) (-1) \oplus \left( \mathbf{ 1,2,1 } \right) (1)
\oplus \left( \mathbf{ 1,2,1 } \right) (-1),\\ \nonumber
\Phi_{c} &\sim & \Phi_a^{\dagger}.
\end{eqnarray}
The resulting spectrum of particle masses implies that the $b$ quantities are 
\begin{eqnarray}
\begin{array}{ccccc}
b_{q_1} = -5, & b_{L_1}  =   -\frac{10}{3}+\frac{7 N_H}{6}, & b_{{\ell}_1}  = -\frac{10}{3}+\frac{N_H}{2}, 
& b_{R_1}  =  -5 +\frac{3 N_H}{2}, & b_{u_1} =  4 + \frac{2 N_H}{3}, \\
b_{q_2}  =  -7,  & b_{L_2}   =  -\frac{10}{3}+\frac{7 N_H}{6}, & b_{{\ell}_2} = -\frac{22}{3}, &
 b_{R_2} = -\frac{10}{3}+\frac{7 N_H}{6},&  b_{u_2}  =  \frac{8}{3}+N_H, \\
b_{q_3}  =  b_{q_2},&  b_{L_3}  =  b_{L_2},& b_{{\ell}_3} =  b_{{\ell}_2},&&
 b_{u_3}  =  \frac{20}{3}+\frac{7 N_H}{6}.
\end{array}
\end{eqnarray}

Substituting these in, the equations describing the evolution of the gauge coupling constants are
\begin{eqnarray}
\centering
\frac{1}{\alpha_q (v)} & = & \frac{1}{\alpha_q} + \frac{5}{2 \pi} \ln{ \left( \frac{v}{M_{EW}} \right) } 
+ \frac{1}{\pi} \ln{ \left( \frac{w}{M_{EW}} \right)}, \\ 
\frac{1}{\alpha_L (v)} & = & \frac{1}{\alpha_L} + \frac{20-7 N_H}{12 \pi} \ln{ \left( \frac{v}{M_{EW}} \right) } ,\\ 
\frac{1}{\alpha_{\ell} (v)} & = & \frac{1}{\alpha_{\ell}} 
+ \frac{20- 3 N_H}{12 \pi} \ln{ \left( \frac{v}{M_{EW}} \right) } 
+ \frac{8 +  N_H }{4 \pi} \ln{ \left( \frac{w}{M_{EW}} \right) } ,\\ 
\frac{1}{\alpha_Y} & = & \frac{1}{3 \alpha_{\ell}} + \frac{5}{3 \alpha_L} 
+ \frac{36+5 N_H}{18 \pi} \ln{ \left( \frac{v}{M_{EW}} \right) }
+ \frac{10- N_H }{6 \pi} \ln{ \left( \frac{w}{M_{EW}} \right) }
+ \frac{22-3 N_H }{6 \pi} \ln{ \left( \frac{x}{M_{EW}} \right) }.
\end{eqnarray}
As a consequence, we have 
\begin{equation}
x_{min} = w \sim 6.5 \times 10^5 GeV, \qquad v \sim 3.9 \times 10^{19} GeV, \qquad \alpha^{-1}_{G_4} = 43.55, 
\qquad \alpha_{\ell} = 0.0526
\end{equation}
as the minimum possible value at which the last breaking can occur. The maximum value of this energy scale is
\begin{equation}
x_{max}  \sim 6.5 \times 10^7 GeV, \qquad w=v \sim 7.4 \times 10^{13} GeV, \qquad  \alpha^{-1}_{G_4} = 39.01, 
\qquad \alpha_{\ell} = 0.1407.
\end{equation}
The GUT unification scale is quite high in this scenario.

\subsection{Cascade 3}

The general form of the equations for this symmetry breaking pattern is given by 
Eq.~\ref{eqn:RGEgeneric} together with
\begin{eqnarray}
\frac{1}{\alpha_Y} & = &
\frac{1}{3 \alpha_{\ell}} + \frac{5}{3 \alpha_L} 
+ \frac{3 (b_{u_1} +  b_{R_1})-5 b_{L_1}}{6 \pi} \ln{\left(\frac{v}{M_{EW}}\right)} \nonumber\\
& + & \frac{3 (b_{u_2} +  b_{R_2}- b_{u_1} -  b_{R_1}) - b_{\ell_2} + 5 (b_{L_1}-
 b_{L_2})}{6 \pi} \ln{\left(\frac{w}{M_{EW}}\right)} \\ \nonumber
& + & \frac{3( b_{u_3} -  b_{u_2} -  b_{R_2}) + b_{\ell_2}- b_{\ell_3}+ 5 (b_{L_2}-
b_{L_3})}{6 \pi} \ln{\left(\frac{x}{M_{EW}}\right)}.
\end{eqnarray}
The light Higgs spectrum has the form 
\begin{eqnarray}
\Phi_{\ell} & \stackrel{v}{\rightarrow} & \left( \mathbf{ 1,2, \overline{3},1 } \right) \left(-\frac{1}{3}\right) 
\oplus  \left( \mathbf{ 1,1,\overline{3},1 } \right) \left(\frac{2}{3}\right)
\stackrel{w}{\rightarrow} \left( \mathbf{ 1,2,1,1 } \right) (-1) 
\stackrel{x}{\rightarrow} \left( \mathbf{ 1,2,1} \right) (-1), \\ \nonumber
\Phi_{\ell^c} & \stackrel{v}{\rightarrow} & \left( \mathbf{ 1,1,3,2 } \right) \left(\frac{1}{3}\right) 
\oplus  \left( \mathbf{ 1,1,\overline{3},1 } \right) \left(-\frac{2}{3}\right)
\stackrel{w}{\rightarrow} \left( \mathbf{ 1,1,1,2 } \right) (1),
\stackrel{x}{\rightarrow} nothing, \\ \nonumber  
\Phi_a & \stackrel{v,w}{\rightarrow} & \left( \mathbf{ 1,2,1,2 } \right) (0) \oplus \left( \mathbf{ 1,1,1,2 } \right) (1)
\oplus \left( \mathbf{ 1,2,1,1 } \right) (-1), \\ \nonumber  
& \stackrel{x}{\rightarrow} & \left( \mathbf{ 1,2,1 } \right) (-1) \oplus \left( \mathbf{ 1,2,1 } \right) (1)
\oplus \left( \mathbf{ 1,2,1 } \right) (-1),\\ \nonumber
\Phi_{c} &\sim &\Phi_a^{\dagger}, 
\end{eqnarray}
specifying the $b$'s as
\begin{eqnarray}
\begin{array}{ccccc}
b_{q_1} = -7, & b_{L_1} = -\frac{4}{3}+\frac{3 N_H}{2}, & b_{{\ell}_1}= -7+N_H, 
& b_{R_1}= -\frac{4}{3}+\frac{3 N_H}{2}, & b_{u_1}=\frac{4}{3} + N_H, \\
b_{q_2} = b_{q_1}, & b_{L_2}=-\frac{10}{3}+\frac{7 N_H}{6}, & b_{{\ell}_2}=-\frac{22}{3}, 
& b_{R_2}= -\frac{10}{3}+\frac{7 N_H}{6}, & b_{u_2}=\frac{8}{3}+N_H, \\
b_{q_3} = b_{q_1}, & b_{L_3}=b_{L_2}, & b_{{\ell}_3}=b_{{\ell}_2}, &
& b_{u_3}=\frac{20}{3}+\frac{7 N_H}{6},
\end{array}
\end{eqnarray}

The renormalisation-group equations reduce to
\begin{eqnarray}
\centering
\frac{1}{\alpha_q (v)} & = & \frac{1}{\alpha_q} + \frac{7}{2 \pi} \ln{ \left( \frac{v}{M_{EW}} \right) },  \\ 
\frac{1}{\alpha_L (v)} & = & \frac{1}{\alpha_L} + \frac{8-9 N_H}{12 \pi} \ln{ \left( \frac{v}{M_{EW}} \right) } 
+ \frac{6+N_H }{6 \pi} \ln{ \left( \frac{w}{M_{EW}} \right) } ,\\ 
\frac{1}{\alpha_{\ell} (v)} & = & \frac{1}{\alpha_{\ell}} + \frac{7- N_H}{2 \pi} \ln{ \left( \frac{v}{M_{EW}} \right) } 
+ \frac{1 + 3 N_H }{6 \pi} \ln{ \left( \frac{w}{M_{EW}} \right) }, \\ 
\frac{1}{\alpha_Y} & = & \frac{1}{3 \alpha_{\ell}} + \frac{5}{3 \alpha_L} 
+ \frac{20}{18 \pi} \ln{ \left( \frac{v}{M_{EW}} \right) }+ \frac{23+ N_H }{9 \pi} 
\ln{ \left( \frac{w}{M_{EW}} \right) }
+ \frac{22-3 N_H }{6 \pi} \ln{ \left( \frac{x}{M_{EW}} \right) }.
\end{eqnarray}
These unify for the range
\begin{eqnarray}
\centering
& x_{min} & \sim 6.3 \times 10^7 GeV, \qquad w=v \sim 7.7 \times 10^{13} GeV, \\ 
& x_{max} &= w \sim 4.9 \times 10^{10} GeV, \qquad v \sim 7 \times 10^{12} GeV.
\end{eqnarray}

\subsection{Cascade 4}
Again the evolution of the $SU(N)$ fine-structure constants is given by 
Eq.~\ref{eqn:RGEgeneric} and the $U(1)$ charge equation has the form 
\begin{eqnarray}
\frac{1}{\alpha_Y} & = &
\frac{1}{3 \alpha_{\ell}} + \frac{5}{3 \alpha_L} 
+ \frac{3 b_{u_1} + 3 b_{R_1}-5 b_{L_1}}{6 \pi} \ln{\left(\frac{v}{M_{EW}}\right)} 
+ \frac{3 b_{u_2} -3 b_{u_1} - 3 b_{R_1} + 5 b_{L_1}-
5 b_{L_2}}{6 \pi} \ln{\left(\frac{w}{M_{EW}}\right)} \\ \nonumber
& + & \frac{3 b_{u_3} - 3 b_{u_2} - b_{\ell_3}+ 5 b_{L_2}-5 b_{L_3}}{6 \pi} \ln{\left(\frac{x}{M_{EW}}\right)}.
\end{eqnarray}
The light Higgs spectrum goes as 
\begin{eqnarray}
\Phi_{\ell} & \stackrel{v,w}{\rightarrow} & \left( \mathbf{ 1,2, \overline{3},1 } \right) \left(-\frac{1}{3}\right) 
\oplus  \left( \mathbf{ 1,1,\overline{3},1 } \right) \left(\frac{2}{3}\right) 
\stackrel{x}{\rightarrow} \left( \mathbf{ 1,2,1} \right) (-1), \\ \nonumber
\Phi_{\ell^c} & \stackrel{v}{\rightarrow} & \left( \mathbf{ 1,1,3,2 } \right) \left(\frac{1}{3}\right) 
\oplus  \left( \mathbf{ 1,1,\overline{3},1 } \right) \left(-\frac{2}{3}\right)
\stackrel{w}{\rightarrow} \left( \mathbf{ 1,1,3 } \right) \left(-\frac{2}{3}\right) 
\oplus  \left( \mathbf{ 1,1,3 } \right) \left(\frac{4}{3}\right) \oplus 
\left( \mathbf{ 1,1,3 } \right) \left(-\frac{2}{3}\right), \\ \nonumber 
\Phi_a & \stackrel{v}{\rightarrow} & \left( \mathbf{ 1,2,1,2 } \right) (0) \oplus \left( \mathbf{ 1,1,1,2 } \right) (1)
\oplus \left( \mathbf{ 1,2,1,1 } \right) (-1), \\ \nonumber  
& \stackrel{w,x}{\rightarrow} & \left( \mathbf{ 1,2,1 } \right) (-1) \oplus \left( \mathbf{ 1,2,1 } \right) (1)
\oplus \left( \mathbf{ 1,2,1 } \right) (-1),\\ \nonumber
\Phi_{c} &\sim &\Phi_a^{\dagger},
\end{eqnarray}
resulting in the quantities
\begin{equation}
\begin{array}{ccccc}
b_{q_1} = -7, & b_{L_1} = -\frac{4}{3}+\frac{3 N_H}{2},&  b_{{\ell}_1}= -7+ N_H, 
& b_{R_1}= -\frac{4}{3}+\frac{3 N_H}{2}, & b_{u_1}=\frac{4}{3} + N_H, \\
b_{q_2} = b_{q_1}, & b_{L_2}=b_{L_1}, & b_{{\ell}_2}=b_{{\ell}_1}, &
& b_{u_2}=\frac{22}{3}+\frac{11 N_H}{6}, \\
b_{q_3} = b_{q_1}, & b_{L_3}=-\frac{10}{3}+\frac{7 N_H}{6}, & b_{{\ell}_3}=-\frac{22}{3},&
 & b_{u_3}=\frac{20}{3}+\frac{7 N_H}{6},
\end{array}
\end{equation}

and the equations
\begin{eqnarray}
\centering
\frac{1}{\alpha_q (v)} & = & \frac{1}{\alpha_q} + \frac{7}{2 \pi} \ln{ \left( \frac{v}{M_{EW}} \right) } , \\ 
\frac{1}{\alpha_L (v)} & = & \frac{1}{\alpha_L} + \frac{8-9 N_H}{12 \pi} \ln{ \left( \frac{v}{M_{EW}} \right) } 
+ \frac{6+N_H }{6 \pi} \ln{ \left( \frac{x}{M_{EW}} \right) }, \\ 
\frac{1}{\alpha_{\ell} (v)} & = & \frac{1}{\alpha_{\ell}} + \frac{7- N_H}{2 \pi} \ln{ \left( \frac{v}{M_{EW}} \right) } 
+ \frac{1 + 3 N_H }{6 \pi} \ln{ \left( \frac{x}{M_{EW}} \right) } ,\\ 
\frac{1}{\alpha_Y} & = & \frac{1}{3 \alpha_{\ell}} + \frac{5}{3 \alpha_L} 
+ \frac{20}{18 \pi} \ln{ \left( \frac{v}{M_{EW}} \right) }+ \frac{11- N_H }{3 \pi} \ln{ \left( \frac{w}{M_{EW}} \right) }
+ \frac{46- N_H }{18 \pi} \ln{ \left( \frac{x}{M_{EW}} \right) }. 
\end{eqnarray}
These unify in a similar range of energy scales to cascade three. 

\section{RGEs for $SU(2)_{\ell}$ broken}\label{cha:appendix2}

We now provide the technical details for the eight cascades featuring completely
broken leptonic colour.

\subsection{Cascade 1}

The VEV pattern that induces the breaking of cascade one is
\begin{equation}
\centering
\langle \Phi_{\ell} \rangle = \left( \begin{array}{ccc}
u & 0 & u \\
0 & u & 0 \\
y & 0 & w \end{array} \right), \qquad 
\langle \Phi_{\ell^c} \rangle = \left( \begin{array}{ccc}
y & 0 & y \\
0 & y & 0 \\
x & 0 & v  \end{array} \right), \qquad
\langle \Phi_{a} \rangle = \langle \Phi_c^{\dagger} \rangle = \left( \begin{array}{ccc}
u & 0 & u \\
0 & u & 0 \\
x & 0 & w  \end{array} \right),
\end{equation}
where $v \geq w \geq x \geq y \geq u$, and $u$ instigates the electroweak symmetry breaking.  
After the first stage of symmetry breaking, the particles $x_1^c, \, x_2^c, \, y_1^c$ and $y_2^c$ 
gain Dirac masses and our light Higgs spectrum is
\begin{eqnarray}
\centering
\Phi_{\ell} & \sim & \left( {\mathbf{1,3,2,1}} \right) \left( \frac{1}{3} \right) \oplus
\left( {\mathbf{1,3,1,1}} \right) \left( -\frac{2}{3} \right), \qquad 
\Phi_{\ell^c} \sim  \left( {\mathbf{1,1,2,2}} \right) \left( 0 \right) \oplus
\left( {\mathbf{1,1,2,1}} \right) \left( -1 \right) 
\oplus \left( {\mathbf{1,1,1,2}} \right) \left( 1 \right), \\ \nonumber
\Phi_a & \sim & \left( {\mathbf{1,3,1,2}} \right) \left( \frac{1}{3} \right) \oplus
\left( {\mathbf{1,3,1,1}} \right) \left( -\frac{2}{3} \right), \qquad
\Phi_c \sim \left( {\mathbf{1,\overline{3},1,2}} \right) \left( -\frac{1}{3} \right) \oplus
\left( {\mathbf{1,\overline{3},1,1}} \right) \left( \frac{2}{3} \right).
\end{eqnarray}
The second stage of breaking sees the remaining charged exotic 
fermions gaining Dirac masses of order $w$, 
and the neutral exotic particle $N,N^c$ gains a $w$ scale Majorana mass. The components of the 
Higgs multiplets which remain light are 
\begin{eqnarray}
\centering
\Phi_{\ell} & \sim & \left( {\mathbf{1,2,2,1}} \right) \left( 0 \right) \oplus
\left( {\mathbf{1,1,2,1}} \right) \left( 1 \right) \oplus 
\left( {\mathbf{1,2,1,1}} \right) \left( -1 \right), \\ \nonumber
\Phi_{\ell^c} &\sim& \left( {\mathbf{1,1,2,2}} \right) \left( 0 \right) \oplus
\left( {\mathbf{1,1,2,1}} \right) \left( -1 \right) \oplus 
\left( {\mathbf{1,1,1,2}} \right) \left( 1 \right), \\ \nonumber
\Phi_a & \sim & \left( {\mathbf{1,2,1,2}} \right) \left( 0 \right) \oplus
\left( {\mathbf{1,1,1,2}} \right) \left( 1 \right) \oplus 
\left( {\mathbf{1,2,1,1}} \right) \left( -1 \right), \\ \nonumber
\Phi_c &\sim &\left( {\mathbf{1,2,1,2}} \right) \left( 0 \right) \oplus
\left( {\mathbf{1,1,1,2}} \right) \left( -1 \right) \oplus \left( {\mathbf{1,2,1,1}} \right) \left( 1 \right).
\end{eqnarray}
There are no fermion mass terms 
of order $x$, but the light Higgs sector reduces to
\begin{eqnarray}
\Phi_{\ell} & \sim & \left( {\mathbf{1,2,2}} \right) \left( 0 \right) \oplus
\left( {\mathbf{1,1,2}} \right) \left( 1 \right) \oplus \left( {\mathbf{1,2,1}} \right) \left( -1 \right), \qquad
\Phi_{\ell^c} \sim \left( \mathbf{ 1,1,2 } \right) \left(-1 \right) \oplus
\left( \mathbf{ 1,1,2 } \right) \left(1 \right) \oplus\left( \mathbf{ 1,1,2 } \right) \left(-1 \right),\\ \nonumber
\Phi_a & \sim & \left( {\mathbf{1,2,1}} \right) \left( -1 \right) \oplus
\left( {\mathbf{1,2,1}} \right) \left( 1 \right) \oplus \left( {\mathbf{1,2,1}} \right) \left( -1 \right), \qquad
\Phi_c \sim \left( {\mathbf{1,2,1}} \right) \left( 1 \right) \oplus
\left( {\mathbf{1,2,1}} \right) \left( -1 \right) \oplus \left( {\mathbf{1,2,1}} \right) \left( 1 \right). 
\end{eqnarray}
After the final stage of breaking down to the standard model gauge group, 
the left-handed anti-neutrino $\nu^c$ gains a $y$ scale mass and 
we have nine light Higgs doublets with $Y= \pm 1$. 
This spectrum of particles defines the $b$ quantities as 
\begin{equation}
\begin{array}{ccccc}
b_{q_1}  = -5 , &  b_{L_1} = -5 + \frac{3 N_H}{2}, 
& b_{\ell_1} = -\frac{10}{3}+N_H, & b_{R_1} = -\frac{10}{3}+\frac{3 N_H}{2}, 
& b_{u_1}  = 4+ \frac{5 N_H}{6},\\
b_{q_2} = -7, & b_{L_2} = -\frac{10}{3}+\frac{3 N_H}{2},& 
b_{\ell_2} =  -\frac{22}{3}+N_H, & b_{R_2} = b_{R_1}, & b_{u_2}  = \frac{8}{3}+\frac{4 N_H}{3},\\
b_{q_3} = b_{q_2}, &   b_{L_3}  = b_{L_2}, & b_{\ell_3} = b_{\ell_2}, 
&& b_{u_3}  =\frac{20}{3}+\frac{11 N_H}{6} ,\\
b_{q_4}  =  b_{q_2},&  b_{L_4}  = b_{L_2},& & &  b_{u_4}  = \frac{20}{3}+\frac{3 N_H}{2}.
\end{array}
\end{equation}
The relationship between the fine-structure constants and the symmetry breaking scales has the general form 
\begin{eqnarray}
\frac{1}{\alpha_Y} & = & \frac{3}{\alpha_L} + \frac{3 (b_{\ell_1} +  b_{R_1}+  b_{u_1})
- 8b_{L_1}}{6 \pi} \ln{ \left( \frac{v}{M_{EW}} \right)} \nonumber\\
& + & \frac{3 \left( b_{\ell_2} + b_{R_2}+ b_{u_2} -b_{\ell_1} - b_{R_1}- b_{u_1} \right) 
+ 8b_{L_1}- 9 b_{L_2}}{6 \pi} \ln{ \left( \frac{w}{M_{EW}} \right)} \nonumber\\
& + &  \frac{ b_{\ell_3} + 3b_{L_2}+ b_{u_3} -b_{\ell_2} - b_{R_2}- b_{u_2} -3 b_{L_3}}{2 \pi} 
\ln{ \left( \frac{x}{M_{EW}} \right)}\nonumber\\ 
& + & \frac{ b_{u_4} -b_{\ell_3} - b_{u_3} +3 b_{L_3} - 3 b_{L_4}}{2 \pi} \ln{ \left( \frac{y}{M_{EW}} \right)}. 
\end{eqnarray}
Inputing these values, the renormalisation-group equations reduce to
\begin{eqnarray}
\frac{1}{\alpha_q (v)} & = & \frac{1}{\alpha_q} + \frac{5}{2 \pi} \ln{ \left( \frac{v}{M_{EW}} \right) }
+ \frac{1}{\pi} \ln{ \left( \frac{w}{M_{EW}} \right) }\\
\frac{1}{\alpha_L (v)} & = & \frac{1}{\alpha_L}+ \frac{10-3N_H}{4 \pi} \ln{ \left( \frac{v}{M_{EW}} \right) } 
- \frac{5}{6 \pi} \ln{ \left( \frac{w}{M_{EW}} \right) }\\
\frac{1}{\alpha_Y} & = & \frac{3}{\alpha_L} + \frac{16-N_H}{3 \pi} \ln{ \left( \frac{v}{M_{EW}} \right) }
- \frac{13}{3 \pi} \ln{ \left( \frac{w}{M_{EW}} \right) }\nonumber\\
& + & \frac{22-3 N_H}{6 \pi}\ln{ \left( \frac{x}{M_{EW}} \right) } 
+ \frac{11-2 N_H}{3 \pi} \ln{ \left( \frac{y}{M_{EW}} \right) }.
\end{eqnarray}
Unification of the coupling constants at $v$ can only be achieved if 
$y_{max} \sim 7.1\times 10^2\ GeV$, with  
the configuration of our energy scales being 
\begin{equation}
\centering
y_{max} =x \sim 7.1\times 10^2\  GeV, \qquad w = v \sim 1.3 \times 10^{13}\ GeV.
\end{equation}
In this symmetry breaking scheme, the unification scale is of order $10^{13}\ GeV$ and does not have 
much scope to change if we want the coupling constants to intersect. 
There now are only two symmetry breaking stages, with the breaking proceeding via
\begin{equation}
\centering
G_4 \rightarrow SU(3)_q \otimes SU(2)_L \otimes SU(2)_{\ell} \otimes SU(2)_R \otimes U(1)_{X_1} 
\rightarrow  SU(3)_q \otimes SU(2)_L \otimes U(1)_Y.
\end{equation} 

\subsection{Cascade 2}

The symmetry breaking of cascade two is generated by Higgs VEVs of the form
\begin{equation}
\centering
\langle \Phi_{\ell} \rangle = \left( \begin{array}{ccc}
u & 0 & u \\
0 & u & 0 \\
x & 0 & w \end{array} \right), \qquad 
\langle \Phi_{\ell^c} \rangle = \left( \begin{array}{ccc}
y & 0 & x \\
0 & y & 0 \\
y & 0 & v  \end{array} \right), \qquad
\langle \Phi_{a} \rangle = \langle \Phi_c^{\dagger} \rangle = \left( \begin{array}{ccc}
u & 0 & u \\
0 & u & 0 \\
y & 0 & w  \end{array} \right).
\end{equation}
The spectrum of fermion masses is identical to cascade one, and the light Higgs fields 
have a similar form but the $b$'s will differ as we have three multiplets which 
transform non-trivially under $SU(3)_R$. This change in the $b$'s is as per:
\begin{equation}
\begin{array}{ccccc}
b_{q_1}  =  -5 , & b_{L_1}  = -5 + \frac{3 N_H}{2}, & b_{\ell_1} = -\frac{10}{3}+N_H, 
& b_{R_1} = -\frac{10}{3}+\frac{3 N_H}{2}, & b_{u_1}  = 4+ \frac{5 N_H}{6},\\
b_{q_2}  =  -7, & b_{L_2} = -\frac{10}{3}+\frac{3 N_H}{2} , &  b_{\ell_2} =  -\frac{22}{3}+N_H, 
& b_{R_2} = -\frac{10}{3}+\frac{3 N_H}{2}, & b_{u_2}  = \frac{8}{3}+\frac{4 N_H}{3},\\
b_{q_3}  =  b_{q_2}, & b_{L_3}  = b_{L_2},& & b_{R_3} = b_{R_2},
&  b_{u_3}  =\frac{8}{3}+\frac{5 N_H}{3} ,\\
b_{q_4}  =  b_{q_2}, & b_{L_4}  = b_{L_2}, &&& b_{u_4}  = \frac{20}{3}+\frac{3 N_H}{2}.
\end{array}
\end{equation}

The evolution of the strong and weak couplings is identical to that of cascade one, however, the 
Abelian-charge fine-structure constant has a different running with energy as evident in its equation
\begin{eqnarray}
\frac{1}{\alpha_Y} &=& \frac{3}{\alpha_L} + \frac{3 b_{\ell_1} + 3 b_{R_1} + 3 b_{u_1}-8b_{L_1}}{6 \pi} 
\ln{ \left( \frac{v}{M_{EW}} \right)}\nonumber\\ 
& + & \frac{3 \left( b_{\ell_2} + b_{R_2}+ b_{u_2} -b_{\ell_1} - b_{R_1}- b_{u_1} \right) + 8b_{L_1}- 9 b_{L_2}}{6 \pi} 
\ln{ \left( \frac{w}{M_{EW}} \right)}\nonumber\\
& + &  \frac{ b_{R_3} + 3b_{L_2}+ b_{u_3} -b_{\ell_2} - b_{R_2}- b_{u_2} -3 b_{L_3}}{2 \pi} 
\ln{ \left( \frac{x}{M_{EW}} \right)}\nonumber\\ 
& + & \frac{ b_{u_4} -b_{R_3} - b_{u_3} +3 b_{L_3} - 3 b_{L_4}}{2 \pi} \ln{ \left( \frac{y}{M_{EW}} \right)} \\
&=& \frac{3}{\alpha_L} + \frac{16-N_H}{3 \pi} \ln{ \left( \frac{v}{M_{EW}} \right) }
- \frac{13}{3 \pi} \ln{ \left( \frac{w}{M_{EW}} \right) }+ 
\frac{11- N_H}{3 \pi}\ln{ \left( \frac{x}{M_{EW}} \right) } 
+ \frac{22-5 N_H}{6 \pi} \ln{ \left( \frac{y}{M_{EW}} \right) }.
\end{eqnarray}
This cascade has an identical range of scales for unification as the previous scheme, the two cascades becoming 
equivalent once the unification scales have been identified. 
 
\subsection{Cascade 3}

The Higgs VEV pattern which induces the breaking of cascade three is
\begin{equation}
\centering
\langle \Phi_{\ell} \rangle = \left( \begin{array}{ccc}
u & 0 & u \\
0 & u & 0 \\
y & 0 & y \end{array} \right), \qquad 
\langle \Phi_{\ell^c} \rangle = \left( \begin{array}{ccc}
w & 0 & w \\
0 & w & 0 \\
w & 0 & v  \end{array} \right), \qquad
\langle \Phi_{a} \rangle = \langle \Phi_c^{\dagger} \rangle = \left( \begin{array}{ccc}
u & 0 & u \\
0 & u & 0 \\
y & 0 & y  \end{array} \right).
\end{equation}
This symmetry 
breaking scheme is the one option that does not allow the unification of the gauge coupling 
constants, with the renormalisation-group equations 
\begin{eqnarray}
\frac{1}{\alpha_q (v)} & = &  \frac{1}{\alpha_q} + \frac{5}{2 \pi} \ln{ \left( \frac{v}{M_{EW}} \right) }
+ \frac{1}{\pi} \ln{ \left( \frac{y}{M_{EW}} \right) }\\
\frac{1}{\alpha_L (v)} & = & \frac{1}{\alpha_L}+ \frac{10-3N_H}{4 \pi} \ln{ \left( \frac{v}{M_{EW}} \right) } 
- \frac{5}{6 \pi} \ln{ \left( \frac{y}{M_{EW}} \right) }\\
\frac{1}{\alpha_Y} & = &  \frac{3}{\alpha_L} + \frac{16-N_H}{3 \pi} \ln{ \left( \frac{v}{M_{EW}} \right) }
+ \frac{16-2 N_H}{3 \pi} \ln{ \left( \frac{w}{M_{EW}} \right) } 
- \frac{14+3 N_H}{6 \pi} \ln{ \left( \frac{y}{M_{EW}} \right) },
\end{eqnarray}
failing to intersect unless $N_H > 1$.

\subsection{Cascade 4}

The Higgs VEV pattern 
\begin{equation}
\centering
\langle \Phi_{\ell} \rangle = \left( \begin{array}{ccc}
u & 0 & u \\
0 & u & 0 \\
y & 0 & w \end{array} \right), \qquad 
\langle \Phi_{\ell^c} \rangle = \left( \begin{array}{ccc}
y & 0 & y \\
0 & y & 0 \\
x & 0 & w  \end{array} \right), \qquad
\langle \Phi_{a} \rangle = \langle \Phi_c^{\dagger} \rangle = \left( \begin{array}{ccc}
u & 0 & u \\
0 & u & 0 \\
x & 0 & v  \end{array} \right),
\end{equation}
instigates the breaking of cascade four. 
After the first stage of breaking the $h$, $h^c$, $z_1$, $z_2$, $N$, $z_1^c$, $z_2^c$ and $N^c$ 
particles gain mass, and the light Higgs spectrum is
\begin{eqnarray}
\centering
\Phi_{\ell} &\sim& \left( \mathbf{ 1,2,\overline{3},1 } \right)\left(-\frac{1}{3}\right) \oplus 
\left( \mathbf{ 1,1,\overline{3},1 } \right)\left(\frac{2}{3}\right), \qquad 
\Phi_{\ell^c} \sim \left( \mathbf{ 1,1,3,2 } \right)\left(\frac{1}{3}\right) \oplus 
\left( \mathbf{ 1,1,3,1 } \right) \left(-\frac{2}{3}\right), \\ 
\Phi_a &\sim& \left( \mathbf{ 1,2,1,2 } \right) (0) \oplus \left( \mathbf{ 1,2,1,1 } \right) (-1)
\oplus \left( \mathbf{ 1,1,1,2 } \right) (1),\label{eqn:phia} \qquad \Phi_c \sim \Phi_a^{\dagger}.
\end{eqnarray}
At $w$, the remaining charged fermions acquire mass, and $\nu^c$ gets an order $y$ mass. 
The components of the Higgs multiplets that remain light are:
\begin{eqnarray}
\centering
\Phi_{\ell} &\sim& \left( \mathbf{ 1,2,2,1 } \right)(0) \oplus 
\left( \mathbf{ 1,1,2,1 } \right)(1) \oplus \left( \mathbf{ 1,2,1,1 } \right) (-1)\label{eqn:phila}, \\  
\Phi_{\ell^c} &\sim& \left( \mathbf{ 1,1,2,2 } \right)(0) \oplus 
\left( \mathbf{ 1,1,2,1 } \right)(-1)\oplus \left( \mathbf{ 1,1,1,2 } \right) (1), \\ 
\Phi_a &\sim & \left( \mathbf{ 1,2,1,2 } \right) (0) \oplus \left( \mathbf{ 1,2,1,1 } \right) (-1)
\oplus \left( \mathbf{ 1,1,1,2 } \right) (1),\qquad \Phi_c \sim \Phi_a^{\dagger}
\end{eqnarray}
at $w$, and 
\begin{eqnarray}
\centering
\Phi_{\ell} &\sim & \left( \mathbf{ 1,2,2 } \right)(0) \oplus 
\left( \mathbf{ 1,1,2 } \right)(1) \oplus \left( \mathbf{ 1,2,1 } \right) (-1), \\  
\Phi_{\ell^c} &\sim &\left( \mathbf{ 1,1,2 } \right)(-1) \oplus 
\left( \mathbf{ 1,1,2 } \right)(-1)\oplus \left( \mathbf{ 1,1,2 } \right) (1), \\ 
\Phi_a & \sim & \left( \mathbf{ 1,2,1 } \right)(1) \oplus 
\left( \mathbf{ 1,2,1 } \right)(1)\oplus \left( \mathbf{ 1,2,1 } \right) (-1),
\qquad \Phi_c \sim \Phi_a^{\dagger}
\end{eqnarray}
at $x$, and at $y$
\begin{eqnarray}
\centering
\Phi_{\ell} &\sim& \left( \mathbf{ 1,2 } \right)(1) \oplus 
\left( \mathbf{ 1,2 } \right)(1)\oplus \left( \mathbf{ 1,2 } \right) (-1),\\
\Phi_a & \sim & \left( \mathbf{ 1,2 } \right)(1) \oplus 
\left( \mathbf{ 1,2 } \right)(1)\oplus \left( \mathbf{ 1,2 } \right) (-1), \qquad \Phi_c \sim \Phi_a^{\dagger}.
\end{eqnarray}
This spectrum of particles defines the $b$'s as follows
\begin{equation}
\begin{array}{ccccc}
b_{q_1} = -7, & b_{L_1}  = -\frac{4}{3}+\frac{3 N_H}{2}, & b_{\ell_1} = -7+N_H, 
& b_{R_1} =  -\frac{4}{3}+\frac{3 N_H}{2}, & b_{u_1}=\frac{4}{3}+ N_H ,\\
b_{q_2} = b_{q_1}, & b_{L_2} =-\frac{10}{3}+\frac{3 N_H}{2}, 
& b_{\ell_2} =  -\frac{22}{3}+N_H, & b_{R_2} =  -\frac{10}{3}+\frac{3 N_H}{2}, 
& b_{u_2}  =  \frac{8}{3}+\frac{4 N_H}{3},\\
b_{q_3} =  b_{q_1}, & b_{L_3} = b_{L_2}, & b_{\ell_3} = b_{\ell_2}, 
&& b_{u_3}  = \frac{20}{3}+\frac{11 N_H}{6},\\
b_{q_4} = b_{q_1}, & b_{L_4} = b_{L_2},&& & b_{u_4}  = \frac{20}{3}+\frac{3 N_H}{2}.
\end{array}
\end{equation}
The evolution of the $U(1)$ factor yields the relation
\begin{eqnarray}
\centering
\frac{1}{\alpha_Y} &=& \frac{3}{\alpha_L} + \frac{ 3b_{u_1}+3 b_{R_1} +4 b_{\ell_1}-9b_{L_1} }{6 \pi} 
\ln{ \left( \frac{v}{M_{EW}} \right)}\nonumber\\ 
& + & \frac{ 3(b_{u_2}+ b_{R_2} + b_{\ell_2} - b_{u_1}-b_{R_1}) -4 b_{\ell_1} +9b_{L_1}
-9b_{L_2} }{6 \pi} \ln{ \left( \frac{w}{M_{EW}} \right) } \nonumber\\
&+& \frac{b_{u_3} + b_{\ell_3}-b_{u_2}- b_{R_2} - b_{\ell_2}+3( b_{L_2}-b_{L_3})}{2 \pi} 
\ln{ \left( \frac{x}{M_{EW}} \right) } \nonumber\\
& + & \frac{b_{u_4} -b_{u_3} - b_{\ell_3}+3( b_{L_3}-b_{L_4})}{2 \pi} \ln{ \left( \frac{y}{M_{EW}} \right) },
\end{eqnarray}
which gives the renormalisation-group equations to be 
\begin{eqnarray}
\frac{1}{\alpha_q (v)} & = & \frac{1}{\alpha_q} + \frac{7}{2 \pi} \ln{ \left( \frac{v}{M_{EW}} \right) },\\
\frac{1}{\alpha_L (v)} & = & \frac{1}{\alpha_L}+ \frac{8-9N_H}{12 \pi} \ln{ \left( \frac{v}{M_{EW}} \right) }
+ \frac{1}{ \pi} \ln{ \left( \frac{w}{M_{EW}} \right) }, \\
\frac{1}{\alpha_Y} & = & \frac{3}{\alpha_L} - \frac{8+N_H}{3 \pi} \ln{ \left( \frac{v}{M_{EW}} \right) }
+ \frac{11}{3 \pi} \ln{ \left( \frac{w}{M_{EW}} \right) }\nonumber\\ 
& + & \frac{22- 3 N_H}{6 \pi} \ln{ \left( \frac{x}{M_{EW}} \right) }
+ \frac{11- 2 N_H}{3 \pi} \ln{ \left( \frac{y}{M_{EW}} \right) }.
\end{eqnarray}
This cascade is less restrictive than the previous three. The 
final breaking stage can occur at the $TeV$ scale, but unlike the first two cascades, this low a value is
not necessary for unification. Again, if we choose $y_{min} \sim 1\ TeV$, then choosing 
$x_{min} \sim y$, yields the maximum scale of unification given by 
$w \sim 6.2 \times 10^{12} GeV$ and $v \sim 1.1 \times 10^{13}\ GeV$, giving the coupling constant 
at $v$ as $\alpha_{G_4}^{-1}=36.82$. If $x$ increases, then both $w$ and $v$ decrease as does $w/v$ until 
we reach $x_{max}=w \sim 4.2 \times 10^7\ GeV$, and $v \sim 3.8 \times 10^{11}\ GeV$. This gives a large 
range of unification possibilities, with the quartification gauge coupling constant equal to 
$\alpha_{G_4}^{-1}=33.12$ at this upper bound. 

The final stage of symmetry breaking can occur up to an energy of $y_{max} \sim 1.2 \times 10^6\ GeV$ while 
still preserving the unification. As $y$ increases, $x_{max}$ decreases as do both $w$ and $v$. At the 
value $y_{max}$, we must have $x=w \sim 1.2 \times 10^6\ GeV$ and $v \sim 1.4 \times 10^{11}\ GeV$ for 
unification, giving the effective coupling $\alpha_{G_4}^{-1}=32.02$. 

\subsection{Cascade 5}

The Higgs VEV pattern which induces the breaking of cascade five is
\begin{equation}
\centering
\langle \Phi_{\ell} \rangle = \left( \begin{array}{ccc}
u & 0 & u \\
0 & u & 0 \\
x & 0 & w \end{array} \right), \qquad 
\langle \Phi_{\ell^c} \rangle = \left( \begin{array}{ccc}
y & 0 & x \\
0 & y & 0 \\
y & 0 & w  \end{array} \right), \qquad
\langle \Phi_{a} \rangle = \langle \Phi_c^{\dagger} \rangle = \left( \begin{array}{ccc}
u & 0 & u \\
0 & u & 0 \\
y & 0 & v  \end{array} \right).
\end{equation}
The fermion mass spectrum is the same as the previous cascade. 
The light Higgs spectrum has the branching
\begin{eqnarray}
\centering
\Phi_{\ell} & \stackrel{v}{\rightarrow} & \left( \mathbf{ 1,2,\overline{3}, 1} \right) \left(-\frac{1}{3}\right) \oplus 
\left( \mathbf{ 1,1,\overline{3}, 1} \right) \left(\frac{2}{3}\right) \\ \nonumber
&\stackrel{w}{\rightarrow}& \left( \mathbf{1,2,2,1} \right) (0) \oplus \left( \mathbf{1,2,1,1} \right) (-1) \oplus 
\left( \mathbf{1,1,2,1} \right) (1) \\ \nonumber
& \stackrel{x}{\rightarrow} & \left( \mathbf{1,2,1} \right) (-1) \oplus
\left( \mathbf{1,2,1} \right) (1)  \oplus \left( \mathbf{1,2,1} \right) (-1)\\ \nonumber
& \stackrel{y}{\rightarrow} & \left( \mathbf{1,2} \right) (-1) \oplus
\left( \mathbf{1,2} \right) (1)  \oplus \left( \mathbf{1,2} \right) (-1)\\
\Phi_{\ell^c} & \stackrel{v}{\rightarrow} & \left( \mathbf{1,1,3,2} \right)\left(\frac{1}{3}\right) \oplus 
\left( \mathbf{1,1,3,1} \right) \left(-\frac{2}{3}\right) \\ \nonumber
&\stackrel{w}{\rightarrow} & \left( \mathbf{1,1,2,2} \right)\left(0\right) \oplus 
\left( \mathbf{1,1,2,1} \right) \left(-1\right) \oplus \left( \mathbf{1,1,1,2} \right) 
\left(1\right) \\ \nonumber 
&\stackrel{x}{\rightarrow} & \left( \mathbf{1,1,2} \right) (-1) \oplus \left( \mathbf{1,1,2} \right) (1) 
\oplus \left( \mathbf{1,1,2} \right) (-1) \stackrel{y}{\rightarrow} nothing \\
\Phi_a & \stackrel{v,w}{\rightarrow} &\left( \mathbf{ 1,2,1,2} \right) (0) \oplus 
\left( \mathbf{ 1,2,1,1} \right) (-1) \oplus \left( \mathbf{ 1,1,1,2} \right) (1)\\ \nonumber
& \stackrel{x}{\rightarrow} & \left( \mathbf{ 1,2,2} \right) (0) \oplus 
\left( \mathbf{ 1,2,1} \right) (-1) \oplus \left( \mathbf{ 1,1,2} \right) (1)\\ \nonumber
& \stackrel{y}{\rightarrow} & \left( \mathbf{ 1,2} \right) (-1) \oplus 
\left( \mathbf{ 1,2} \right) (1) \oplus \left( \mathbf{ 1,2} \right) (-1)\\
\Phi_c & \sim & \Phi_a^{\dagger}.
\end{eqnarray}

The general form for the relationship between the structure constants and the breaking scales is
\begin{eqnarray}
\centering
\frac{1}{\alpha_Y} &=& \frac{3}{\alpha_L} + \frac{ 3b_{u_1}+3 b_{R_1} +4 b_{\ell_1}-9b_{L_1} }{6 \pi} 
\ln{ \left( \frac{v}{M_{EW}} \right)}\nonumber\\ 
& + & \frac{ 3(b_{u_2}+ b_{R_2} + b_{\ell_2} - b_{u_1}-b_{R_1}) -4 b_{\ell_1} +9b_{L_1}
-9b_{L_2} }{6 \pi} \ln{ \left( \frac{w}{M_{EW}} \right) } \nonumber\\
&+& \frac{b_{u_3} + b_{R_3}-b_{u_2}- b_{R_2} - b_{\ell_2}+3( b_{L_2}-b_{L_3})}{2 \pi} 
\ln{ \left( \frac{x}{M_{EW}} \right) }\nonumber\\ 
& + & \frac{b_{u_4} -b_{u_3} - b_{R_3}+3( b_{L_3}-b_{L_4})}{2 \pi} \ln{ \left( \frac{y}{M_{EW}} \right) },
\end{eqnarray}
where the $b$'s are defined as
\begin{equation}
\begin{array}{ccccc}
b_{q_1} = -7, & b_{L_1}  = -\frac{4}{3}+\frac{3 N_H}{2}, & b_{\ell_1} = -7+N_H, 
& b_{R_1} =  -\frac{4}{3}+\frac{3 N_H}{2}, & b_{u_1}=\frac{4}{3}+ N_H ,\\
b_{q_2} = b_{q_1}, & b_{L_2} =-\frac{10}{3}+\frac{3 N_H}{2} , 
& b_{\ell_2} =  -\frac{22}{3}+N_H, & b_{R_2} =  -\frac{10}{3}+\frac{3 N_H}{2}, 
& b_{u_2}  =  \frac{8}{3}+\frac{4 N_H}{3},\\
b_{q_3} = b_{q_1}, & b_{L_3} =b_{L_1} , && b_{R_3}=b_{R_2}, 
& b_{u_3}  = \frac{8}{3}+\frac{5 N_H}{3},\\
b_{q_4} = b_{q_1}, & b_{L_4}  =b_{L_1} , &&& b_{u_4}  = \frac{20}{3}+\frac{3 N_H}{2}.
\end{array}
\end{equation}
The renormalisation-group equations are
\begin{eqnarray}
\frac{1}{\alpha_q (v)} & = & \frac{1}{\alpha_q} + \frac{7}{2 \pi} \ln{ \left( \frac{v}{M_{EW}} \right) },\\
\frac{1}{\alpha_L (v)} & = & \frac{1}{\alpha_L}+ \frac{8-9N_H}{12 \pi} \ln{ \left( \frac{v}{M_{EW}} \right) }
+ \frac{1}{ \pi} \ln{ \left( \frac{w}{M_{EW}} \right) }, \\
\frac{1}{\alpha_Y} & = & \frac{3}{\alpha_L} - \frac{8+N_H}{3 \pi} \ln{ \left( \frac{v}{M_{EW}} \right) }
+ \frac{11}{3 \pi} \ln{ \left( \frac{w}{M_{EW}} \right) } 
+ \frac{11 - N_H}{3 \pi} \ln{ \left( \frac{x}{M_{EW}} \right) }\nonumber\\
& + & \frac{22- 5 N_H}{6 \pi} \ln{ \left( \frac{y}{M_{EW}} \right) },
\end{eqnarray}
where only the last equation is different from those of the previous cascade. This difference is compensated in the 
$b$'s and we obtain a very similar spectrum of energy scales to cascade four which yield unification.
 
\subsection{Cascade 6}

The VEV pattern that induces the breaking in cascade six is
\begin{equation}
\centering
\langle \Phi_{\ell} \rangle = \left( \begin{array}{ccc}
u & 0 & u \\
0 & u & 0 \\
y & 0 & x \end{array} \right), \qquad 
\langle \Phi_{\ell^c} \rangle = \left( \begin{array}{ccc}
y & 0 & y \\
0 & y & 0 \\
x & 0 & x  \end{array} \right), \qquad
\langle \Phi_{a} \rangle = \langle \Phi_c^{\dagger} \rangle = \left( \begin{array}{ccc}
u & 0 & u \\
0 & u & 0 \\
w & 0 & v  \end{array} \right).
\end{equation}
The particles $h$, $h^c$, $z_1$, $z_2$, $N$, $z_1^c$, $z_2^c$ and $N^c$ gain order $v$ masses, while 
there are no new masses at $w$. At $x$ the remaining exotic fermions acquire mass, and as 
usual, $\nu^c$ has mass of order $y$.
The Higgs spectrum which is light has the branching
\begin{eqnarray}
\centering
\Phi_{\ell} & \stackrel{v,w}{\rightarrow} & \left( \mathbf{ 1,2,\overline{3}, 1} \right) \left(-\frac{1}{3}\right) \oplus 
\left( \mathbf{ 1,1,\overline{3}, 1} \right) \left(\frac{2}{3}\right) \\ \nonumber
&\stackrel{x}{\rightarrow}& \left( \mathbf{1,2,2} \right) (0) \oplus \left( \mathbf{1,2,1} \right) (-1) \oplus 
\left( \mathbf{1,1,2} \right) (1) \\ \nonumber
& \stackrel{y}{\rightarrow} & \left( \mathbf{1,2} \right) (-1) \oplus
\left( \mathbf{1,2} \right) (1)  \oplus \left( \mathbf{1,2} \right) (-1)\\
\Phi_{\ell^c} & \stackrel{v}{\rightarrow} & \left( \mathbf{1,1,3,2} \right)\left(\frac{1}{3}\right) \oplus 
\left( \mathbf{1,1,3,1} \right) \left(-\frac{2}{3}\right) \\ \nonumber
&\stackrel{w}{\rightarrow} & \left( \mathbf{1,1,3} \right)\left(-\frac{2}{3}\right) \oplus 
\left( \mathbf{1,1,3} \right) \left(\frac{4}{3}\right) \oplus \left( \mathbf{1,1,3} \right) 
\left(-\frac{2}{3}\right) \\ \nonumber 
&\stackrel{x}{\rightarrow} & \left( \mathbf{1,1,2} \right) (-1) \oplus \left( \mathbf{1,1,2} \right) (1) 
\oplus \left( \mathbf{1,1,2} \right) (-1) \stackrel{y}{\rightarrow} nothing \\
\Phi_a & \stackrel{v}{\rightarrow} &\left( \mathbf{ 1,2,1,2} \right) (0) \oplus 
\left( \mathbf{ 1,2,1,1} \right) (-1) \oplus \left( \mathbf{ 1,1,1,2} \right) (1)\\ \nonumber
& \stackrel{w,x,y}{\rightarrow} & \left( \mathbf{ 1,2,1} \right) (-1) \oplus 
\left( \mathbf{ 1,2,1} \right) (1) \oplus \left( \mathbf{ 1,2,1} \right) (-1)\\
\Phi_c & \sim & \Phi_a^{\dagger}.
\end{eqnarray}
The relationship between the fine-structure constants and the breaking scales is 
\begin{eqnarray}
\centering
\frac{1}{\alpha_Y} &=& \frac{3}{\alpha_L} + \frac{ 3b_{u_1}+3 b_{R_1} +4 b_{\ell_1}-9b_{L_1} }{6 \pi} 
\ln{ \left( \frac{v}{M_{EW}} \right)}\nonumber\\ 
& + & \frac{ 3(b_{u_2} - b_{u_1}-b_{R_1}) + 4 (b_{\ell_2}-b_{\ell_1}) +9b_{L_1}
-9b_{L_2} }{6 \pi} \ln{ \left( \frac{w}{M_{EW}} \right) } \nonumber\\
&+& \frac{3(b_{u_3} + b_{\ell_3}-b_{u_2}) -4 b_{\ell_2}+9( b_{L_2}-b_{L_3})}{6 \pi} 
\ln{ \left( \frac{x}{M_{EW}} \right) } \nonumber\\
& + & \frac{b_{u_4} -b_{u_3} - b_{\ell_3}+3( b_{L_3}-b_{L_4})}{2 \pi} \ln{ \left( \frac{y}{M_{EW}} \right) },
\end{eqnarray}
with the $b$'s defined by
\begin{equation}
\begin{array}{ccccc}
b_{q_1}  =  -7,&  b_{L_1}  = -\frac{4}{3}+\frac{3 N_H}{2}, &  b_{\ell_1} = -7+N_H, &
b_{R_1} =  -\frac{4}{3}+\frac{3 N_H}{2}, & b_{u_1}=\frac{4}{3}+ N_H \\
b_{q_2}  =  b_{q_1}, & b_{L_2}  = -\frac{4}{3}+\frac{3 N_H}{2},& b_{\ell_2} = b_{\ell_1}, &&
b_{u_2}  = \frac{22}{3}+\frac{11 N_H}{6} \\
b_{q_3}  =  b_{q_1}, &  b_{L_3} = -\frac{10}{3}+\frac{3 N_H}{2},
& b_{\ell_3} = -\frac{22}{3}+N_H,&& b_{u_3}  = \frac{20}{3}+\frac{11 N_H}{6},\\
b_{q_4}  =  b_{q_1},& b_{L_4}  = b_{L_3},&&& b_{u_4}  = \frac{20}{3}+\frac{3 N_H}{2}.
\end{array}
\end{equation}
The equations reduce to 
\begin{eqnarray}
\frac{1}{\alpha_q (v)} & = & \frac{1}{\alpha_q} + \frac{7}{2 \pi} \ln{ \left( \frac{v}{M_{EW}} \right) },\\
\frac{1}{\alpha_L (v)} & = & \frac{1}{\alpha_L}+ \frac{8-9N_H}{12 \pi} \ln{ \left( \frac{v}{M_{EW}} \right) }
+ \frac{1}{ \pi} \ln{ \left( \frac{x}{M_{EW}} \right) } ,\\
\frac{1}{\alpha_Y} & = & \frac{3}{\alpha_L} - \frac{8+ N_H}{3 \pi} \ln{ \left( \frac{v}{M_{EW}} \right) }
+ \frac{11-N_H}{3 \pi} \ln{ \left( \frac{w}{M_{EW}} \right) }\nonumber\\ 
& + & \frac{22- N_H}{6 \pi} \ln{ \left( \frac{x}{M_{EW}} \right) }
+ \frac{11- 2 N_H}{3 \pi} \ln{ \left( \frac{y}{M_{EW}} \right) }.
\end{eqnarray}
Again unification can be achieved for a range of values for the lower breaking scales.

\subsection{Cascade 7}

The VEV pattern that induces the breaking pattern of cascade seven is
\begin{equation}
\centering
\langle \Phi_{\ell} \rangle = \left( \begin{array}{ccc}
u & 0 & u \\
0 & u & 0 \\
y & 0 & v \end{array} \right), \qquad 
\langle \Phi_{\ell^c} \rangle = \left( \begin{array}{ccc}
y & 0 & x \\
0 & y & 0 \\
y & 0 & w  \end{array} \right), \qquad
\langle \Phi_{a} \rangle = \langle \Phi_c^{\dagger} \rangle = \left( \begin{array}{ccc}
u & 0 & u \\
0 & u & 0 \\
x & 0 & w  \end{array} \right).
\end{equation}
After the first stage of symmetry breaking the exotic fermions $x_1$, $x_2$, $y_1$ and $y_2$ 
gain GUT scale Dirac masses. The remaining exotic fermions gain $w$ scale masses and 
$\nu^c$ an order $y$ mass.
The light Higgs spectrum has the branching
\begin{eqnarray}
\centering
\Phi_{\ell} & \stackrel{v,w,x}{\rightarrow} & \left( \mathbf{ 1,2,2,1} \right) (0) \oplus 
\left( \mathbf{ 1,2,1,1} \right) (-1) \oplus \left( \mathbf{ 1,1,2,1} \right) (1)\\ \nonumber
& \stackrel{y}{\rightarrow} & \left( \mathbf{1,2} \right) (-1) \oplus
\left( \mathbf{1,2} \right) (1)  \oplus \left( \mathbf{1,2} \right) (-1)\\
\Phi_{\ell^c} & \stackrel{v}{\rightarrow} & \left( \mathbf{1,1,2,\overline{3}} \right)\left(-\frac{1}{3}\right) \oplus 
\left( \mathbf{1,1,1,\overline{3}} \right) \left(\frac{2}{3}\right) \\ \nonumber
& \stackrel{w}{\rightarrow} & \left( \mathbf{ 1,1,2,2} \right) (0) \oplus 
\left( \mathbf{ 1,1,2,1} \right) (-1) \oplus \left( \mathbf{ 1,1,1,2} \right) (1)\\ \nonumber
\Phi_a & \stackrel{v}{\rightarrow} & \left( \mathbf{ 1,2,\overline{3}, 1} \right) \left(-\frac{1}{3}\right) \oplus 
\left( \mathbf{ 1,1,\overline{3}, 1} \right) \left(\frac{2}{3}\right) \\ \nonumber
& \stackrel{w}{\rightarrow} &\left( \mathbf{ 1,2,1,2} \right) (0) \oplus 
\left( \mathbf{ 1,2,1,1} \right) (-1) \oplus \left( \mathbf{ 1,1,1,2} \right) (1)\\ \nonumber
& \stackrel{x,y}{\rightarrow} & \left( \mathbf{ 1,2,1} \right) (-1) \oplus 
\left( \mathbf{ 1,2,1} \right) (1) \oplus \left( \mathbf{ 1,2,1} \right) (-1)\\
\Phi_c & \sim & \Phi_a^{\dagger}.
\end{eqnarray}

The $b$'s are then
\begin{equation}
\begin{array}{ccccc}
b_{q_1} = -5, & b_{L_1} = -\frac{10}{3} +\frac{3N_H}{2}, & b_{\ell_1} = -\frac{10}{3} +N_H,
& b_{R_1} = -5+\frac{3N_H}{2},&  b_{u_1}  = 4+\frac{5N_H}{6},\\
b_{q_2} = -7, & b_{L_2} =b_{L_1} , & b_{\ell_2} = -\frac{22}{3} +N_H, 
& b_{R_2} = -\frac{10}{3} +\frac{3N_H}{2}, & b_{u_2}  = \frac{8}{3}+\frac{4 N_H}{3},\\
b_{q_3} = b_{q_2}, & b_{L_3} = b_{L_1}, & b_{\ell_3} = b_{\ell_2}, 
&& b_{u_3} =\frac{20}{3}+\frac{11 N_H}{6} ,\\
b_{q_4} = b_{q_2}, & b_{L_4} = b_{L_1}, &&& b_{u_4}  = \frac{20}{3}+\frac{3 N_H}{2}.
\end{array}
\end{equation}
The fine-structure constants at the electroweak level are related to the breaking scales via
\begin{eqnarray}
\centering
\frac{1}{\alpha_Y} &=& \frac{3}{\alpha_L} + \frac{ 3b_{u_1}+4 b_{R_1} +3 b_{\ell_1}-9b_{L_1} }{6 \pi} 
\ln{ \left( \frac{v}{M_{EW}} \right)} \nonumber\\
& +  & \frac{ 3(b_{u_2}+ b_{R_2} + b_{\ell_2} - b_{u_1}-b_{\ell_1}) -4 b_{R_1} +9b_{L_1}
-9b_{L_2} }{6 \pi} \ln{ \left( \frac{w}{M_{EW}} \right) } \nonumber\\
&+& \frac{b_{u_3} + b_{\ell_3}-b_{u_2}- b_{R_2} - b_{\ell_2}+3( b_{L_2}-b_{L_3})}{2 \pi} 
\ln{ \left( \frac{x}{M_{EW}} \right) } \nonumber\\
& + & \frac{b_{u_4} -b_{u_3} - b_{\ell_3}+3( b_{L_3}-b_{L_4})}{2 \pi} \ln{ \left( \frac{y}{M_{EW}} \right) },
\end{eqnarray}
giving renormalisation-group equations of the form
\begin{eqnarray}
\frac{1}{\alpha_q (v)} & = & \frac{1}{\alpha_q} + \frac{5}{2 \pi} \ln{ \left( \frac{v}{M_{EW}} \right) }
+ \frac{1}{ \pi} \ln{ \left( \frac{w}{M_{EW}} \right) },\\
\frac{1}{\alpha_L (v)} & = & \frac{1}{\alpha_L}+ \frac{20-9N_H}{12 \pi} \ln{ \left( \frac{v}{M_{EW}} \right) },\\
\frac{1}{\alpha_Y} & = & \frac{3}{\alpha_L} + \frac{6-N_H}{3 \pi} \ln{ \left( \frac{v}{M_{EW}} \right) }
- \frac{1}{ \pi} \ln{ \left( \frac{w}{M_{EW}} \right) }\nonumber\\ 
& + & \frac{22- 3 N_H}{6 \pi} \ln{ \left( \frac{x}{M_{EW}} \right) }
+ \frac{11- 2 N_H}{3 \pi} \ln{ \left( \frac{y}{M_{EW}} \right) }.
\end{eqnarray}

The GUT scale must be of order $10^{13}\ GeV$ for unification. The lowest two 
breaking scales are forced to be around the $TeV$ scale, with the only freedom coming into the 
choice of the $w$ scale. The range of scales for which unification can be achieved are the same 
as cascades one and two, with the symmetry breaking patterns becoming identical. 

\subsection{Cascade 8}

As previously noted, choosing whether or not to break $SU(2)_{\ell}$ or $SU(2)_R$ first from 
the $SU(3)_q \otimes SU(2)_L \otimes SU(2)_{\ell} \otimes SU(2)_R \otimes U(1)$ level has no 
significant difference on the unification scales and our fermion mass spectrum is identical 
to that above. Subsequently, we just list the equations below. 

The VEV pattern that induces the breaking pattern of cascade eight is
\begin{equation}
\centering
\langle \Phi_{\ell} \rangle = \left( \begin{array}{ccc}
u & 0 & u \\
0 & u & 0 \\
x & 0 & v \end{array} \right), \qquad 
\langle \Phi_{\ell^c} \rangle = \left( \begin{array}{ccc}
y & 0 & x \\
0 & y & 0 \\
y & 0 & w  \end{array} \right), \qquad
\langle \Phi_{a} \rangle = \langle \Phi_c^{\dagger} \rangle = \left( \begin{array}{ccc}
u & 0 & u \\
0 & u & 0 \\
y & 0 & w  \end{array} \right).
\end{equation}
The general form of the relationship between the fine-structure constants at low energy and the breaking scales is 
\begin{eqnarray}
\centering
\frac{1}{\alpha_Y} &=& \frac{3}{\alpha_L} + \frac{ 3b_{u_1}+4 b_{R_1} +3 b_{\ell_1}-9b_{L_1} }{6 \pi} 
\ln{ \left( \frac{v}{M_{EW}} \right)} \nonumber\\
& + & \frac{ 3(b_{u_2}+ b_{R_2} + b_{\ell_2} - b_{u_1}-b_{\ell_1}) -4 b_{R_1} +9b_{L_1}
-9b_{L_2} }{6 \pi} \ln{ \left( \frac{w}{M_{EW}} \right) } \nonumber\\
&+& \frac{b_{u_3} + b_{R_3}-b_{u_2}- b_{R_2} - b_{\ell_2}+3( b_{L_2}-b_{L_3})}{2 \pi} 
\ln{ \left( \frac{x}{M_{EW}} \right) }\nonumber\\ 
& + & \frac{b_{u_4} -b_{u_3} - b_{R_3}+3( b_{L_3}-b_{L_4})}{2 \pi} \ln{ \left( \frac{y}{M_{EW}} \right) }.
\end{eqnarray}
The $b$'s are defined by 
\begin{equation}
\begin{array}{ccccc}
b_{q_1} = -5, &b_{L_1} = -\frac{10}{3} +\frac{3N_H}{2}, & b_{\ell_1} = -\frac{10}{3} +N_H, 
& b_{R_1} = -5+\frac{3N_H}{2}, & b_{u_1}  = 4+\frac{5N_H}{6},\\
b_{q_2} = -7, & b_{L_2} = b_{L_1}, & b_{\ell_2} = -\frac{22}{3} +N_H, 
& b_{R_2} = -\frac{10}{3} +\frac{3N_H}{2}, & b_{u_2}  = \frac{8}{3} +\frac{4N_H}{3},\\
b_{q_3} = b_{q_2}, & b_{L_3} = b_{L_1}, & &b_{R_3}=b_{R_2},& b_{u_3}  = \frac{8}{3} +\frac{5N_H}{3},\\
b_{q_4} = b_{q_2}, & b_{L_4} =b_{L_1} ,&& & b_{u_4}  = \frac{20}{3} +\frac{3N_H}{2},
\end{array}
\end{equation}
and the RGEs
\begin{eqnarray}
\frac{1}{\alpha_q (v)} & = & \frac{1}{\alpha_q} + \frac{5}{2 \pi} \ln{ \left( \frac{v}{M_{EW}} \right) }
+ \frac{1}{ \pi} \ln{ \left( \frac{w}{M_{EW}} \right) },\\
\frac{1}{\alpha_L (v)} & = & \frac{1}{\alpha_L}+ \frac{20-9N_H}{12 \pi} \ln{ \left( \frac{v}{M_{EW}} \right) },\\
\frac{1}{\alpha_Y} & = & \frac{3}{\alpha_L} + \frac{6-N_H}{3 \pi} \ln{ \left( \frac{v}{M_{EW}} \right) }
- \frac{1}{ \pi} \ln{ \left( \frac{w}{M_{EW}} \right) } \nonumber\\
& + & \frac{11- N_H}{3 \pi} \ln{ \left( \frac{x}{M_{EW}} \right) }
+\frac{22- 5 N_H}{6 \pi} \ln{ \left( \frac{y}{M_{EW}} \right) }.
\end{eqnarray}

\end{document}